\documentclass{aa}
\usepackage{graphicx}
\usepackage{txfonts}
\usepackage{amsmath}
\usepackage{natbib}
\bibpunct{(}{)}{;}{a}{}{,} 
\usepackage{graphicx}
\usepackage{longtable}
\usepackage{supertabular}
\begin{document}

\title{Baade's window with APOGEE}
\subtitle{Metallicities, ages and  chemical abundances}
\author{M. Schultheis\inst{1}
\and A.~Rojas-Arriagada \inst{1}
\and A.~E.~Garc{\'i}a P\'erez \inst{2,3}
\and H.~J\"onsson \inst{2,3}
\and M.~Hayden \inst{1}
\and G.~Nandakumar \inst{1}
\and K.~Cunha \inst{4}
\and C.~Allende Prieto \inst{2,3}
\and J.~A.~Holtzman \inst{5}
\and T.~C.~Beers \inst{6}
\and D.~Bizyaev \inst{7,8}
\and J.~Brinkmann \inst{7}
\and R.~Carrera \inst{2}
\and R.~E.~Cohen \inst{9}
\and D.~Geisler \inst{9}
\and F.~R.~Hearty \inst{10}
\and J.~G.~ Fern\'andez-Trincado \inst{11}
\and C.~Maraston \inst{12}
\and D.~Minniti \inst{13,14,15}
\and C.~Nitschelm \inst{16}
\and A.~Roman-Lopes \inst{17}
\and D.~P.~Schneider \inst{10,18}
\and B.~Tang \inst{9}
\and S.~Villanova \inst{9}
\and G.~Zasowski \inst{19,20}
\and S.~R.~Majewski \inst{21}
}

 \institute{ Laboratoire Lagrange, Universit\'e C\^ote d'Azur, Observatoire de la C\^ote d'Azur, CNRS, Blvd de l'Observatoire, F-06304 Nice, France
 e-mail: mathias.schultheis@oca.eu
  \and
Instituto de Astrof\'{i}sica de Canarias (IAC), E-38205 La Laguna, Tenerife, Spain
\and
Universidad de La Laguna, Dpto. Astrof\'{i}sica, E-38206 La Laguna, Tenerife, Spain
\and
Observat'orio Nacional, Sao Crist\'ovao, Rio de Janeiro, Brazil
\and
New Mexico State University, Las Cruces, NM 88003, USA
\and
Department of Physics and JINA Center for the Evolution of the Elements, University of Notre Dame, Notre Dame, IN 46556, USA
\and
Apache Point Observatory and New Mexico State University, P.O. Box 59, Sunspot, NM, 88349-0059, USA
\and
Sternberg Astronomical Institute, Moscow State University, Moscow
\and
Departamento de Astronom\'ia, Casilla 160-C, Universidad de Concepci\'on, Concepci\'on, Chile
\and
Department of Astronomy and Astrophysics, The Pennsylvania State University, University Park, PA 16802
\and 
The Observatoire des sciences de l'Univers de Besancon, 41 Avenue de l'Observatoire, 25000 Besancon, France
\and 
ICG-University of Portsmouth, Dennis Sciama Building, Burnaby Rd., Portsmouth - PO13FX - UK
\and
Departamento de Fisica, Facultad de Ciencias Exactas, Universidad Andres Bello Av. Fernandez Concha 700, Las Condes, Santi-
ago, Chile
\and
Millennium Institute of Astrophysics, Av. Vicuna Mackenna 4860, 782-0436 Macul, Santiago, Chile
\and
Vatican Observatory, V00120 Vatican City State, Italy
\and 
Unidad de Astronom\'ia, Universidad de Antofagasta, Avenida Angamos 601, Antofagasta 1270300, Chile
\and
Departamento de Fısica, Facultad de Ciencias, Universidad de La Serena, Cisternas 1200, La Serena, Chile
\and
Institute for Gravitation and the Cosmos, The Pennsylvania State University, University Park, PA 16802
\and
Center for Astrophysical Sciences, Department of Physics and Astronomy, Johns Hopkins University, 3400 North Charles Street, Baltimore, MD 21218, USA
\and
Department of Physics \& Astronomy, University of Utah, 115 S. 1400 E., Salt Lake City, UT 84112, USA
\and
Department of Astronomy, University of Virginia, Charlottesville, VA, 22904, USA
}

\abstract
{Baade's window (BW) is one of the most observed Galactic bulge fields in terms of chemical abundances. Due to its low and homogeneous interstellar absorption it is considered as a calibration field for Galactic bulge studies.}
{In the era of large spectroscopic surveys,  calibration fields such as BW are necessary to cross calibrate the stellar parameters and individual abundances of the APOGEE survey. }
{We use the APOGEE BW stars to  derive their metallicity distribution function (MDF) and  individual abundances, for $\alpha$- and iron-peak elements of the APOGEE ASPCAP pipeline (DR13), as well as  the age distribution for stars in BW. }
{We determine the MDF of APOGEE stars in BW and find a remarkable agreement with that of the Gaia-ESO survey (GES). Both exhibit a clear bimodal distribution.  We also find  that the Mg-metallicity planes of  both surveys agree  well, except for the metal-rich part ($\rm [Fe/H] >0.1$),  where APOGEE finds systematically higher Mg abundances with respect to the GES. The ages based on the $\rm [C/N]$ ratio reveal a bimodal age distribution, with a major old population  at $\rm \sim 10\,Gyr$, with a decreasing tail towards younger stars.  A comparison between APOGEE estimates and stellar parameters, and those  determined by other sources reveals detectable systematic offsets, in particular for spectroscopic surface gravity estimates. In general, we find a good agreement  between individual abundances of O, Na, Mg, Al, Si, K, Ca, Cr, Mn, Co, and Ni from APOGEE with that of literature values.  }
{ We have shown that in general APOGEE data show a good agreement in terms of MDF and individual chemical abundances with respect to literature works. Using the [C/N] ration we found a significant fraction of young stars in BW  which is in agreement with the  model of Haywood et al. (2016).}

\maketitle

\titlerunning{Baade's window with APOGEE }
\authorrunning{Schultheis et al.}

\section{Introduction}
Large high-resolution spectroscopic surveys in the Galactic bulge such as APOGEE \citep{majewski2015}, Gaia-ESO (GES, \citealt{gilmore2010}), or ARGOS (\citealt{ness2013}) may suffer from systematic biases in terms of stellar parameters or individual abundances. In the case of APOGEE, Baade's window (BW) was chosen to validate stellar parameters and chemical abundances with known previous measurements in the literature for high metallicities. Due to its low and homogeneous interstellar extinction (see, e.g.,  \citealt{stanek1996}, \citealt{gonzalez2011}), BW is an ideal field in the Galactic bulge for validation purposes. It also has the advantage of covering a wide range in metallicities.

The era of chemical-abundance studies of the Galactic bulge in BW started in the late 1980s with  \citet{rich1988},  who  obtained metallicities and $\alpha$-elements for 88 K giants. \citet{rich1988}  found evidence for super metal-rich stars and derived a mean $\rm [Fe/H] = +0.2$. \citet{rich1994} acquired high-resolution spectra ($\rm R \sim 17\,000$) of 11 bulge K giants, and revised the abundance distribution in BW with a mean $\rm [Fe/H] = -0.25$, and confirmed enhancements of the $\alpha$-elements Mg and Ti, as well as of Al.  From a much larger sample of K and M giants, \citet{sadler1996} obtained  a similar mean metallicity as \citet{rich1994},  but less enhanced in Mg. Adopting new abundance determination techniques and linelists, tailored to analyze metal-rich giants, \citet{fulbright2006} analyzed 27 K giants, and reported a mean metallicity comparable to that of the local disk. In addition, they found that the $\alpha$-elements O, Mg, Si, Ca, and Ti,  and  the odd-Z  elements Al and Na, are enhanced  relative to the Galactic thin disc, indicating that massive stars contributed to the chemical enrichment of the bulge (\citealt{fulbright2007}). \citet{zoccali2006}  found, based on high-resolution UVES (ESO-VLT) spectra ($\rm R = 47\,000$), enhanced O, Mg, and Al abundances. (see also \citealt{lecureur07}). Contrary to these studies, \citet{melendez08} did not identify any significant difference in the abundance patterns of C, N, and O between the thick disc and the Galactic bulge, suggesting that the bulge and the local thick disc experienced similar chemical evolution histories. \citet{alves-brito10} confirmed this similarity by observing a Galactic bulge sample and a local thick disc sample.

With the advent of the FLAMES multiobject spectrograph at the VLT  (\citealt{pasquini03}), it  became possible to observe a large number of objects simultaneously, at high spectral resolution, a sizable leap for this type of study. \citet{hill2011} analyzed approximately 200  FLAMES/GIRAFFE spectra ($\rm R=20\,000$) of red clump stars in BW, which  revealed the presence of  two populations in their metallicity distribution function (MDF):  a metal-poor component centered on  $\rm [Fe/H] = -0.30$ and $\rm [Mg/H] = -0.06$, with a large dispersion, and a narrow metal-rich component centered around  $\rm [Fe/H] = +0.32$ and  $\rm [Mg/H] =  + 0.35$. These two components also exhibit  kinematical differences, with the metal-poor component compatible with an old spheroid, whereas the metal-rich component is consistent with a population supporting a bar (\citealt{babusiaux2010}). \citet{gonzalez11a} analyzed abundances of Mg, Si, Ca, and Ti for the \citet{zoccali08} sample, and concluded from their large sample of bulge stars that there is  chemical similarity between the thick disc and the bulge. GES observed several bulge fields including BW (\citealt{rojas2014}), and they confirmed the bimodal nature of the MDF using Gaussian-mixture models (GMM). \citet{zoccali16} also obtained the bimodal MDF for  a large number of bulge fields using the GIBS survey. In addition, they found  that these two components have a different spatial distribution, with the metal-poor population  being more centrally concentrated, while the metal-rich population exhibits  a boxy concentration, as expected for a bar seen edge on.

All of the studies mentioned above were based on spectroscopy in the visible wavelength range. With the development of near-IR detectors,  Galactic bulge studies could  begin to take  advantage of the fact that the flux in the near-IR band is several magnitudes higher  than in  the visible for K/M giants,  and that the extinction in the infrared is much less severe (by roughly a factor of 10 in the K-band) compared to the visual.  \citet{rich2005} performed the first detailed abundance analysis of 14 M giants in BW, based on $\rm R=25\,000$ infrared spectra ($\rm 1.5-1.8\, \mu$m) using the NIRSPEC facility at the Keck telescope. They found very similar  iron abundances to those of K giants, but $\alpha$ enhancements  compared to a local disk sample of M giants. \citet{cunha06} determined  individual elemental-abundance estimates  such as C, N, O, Na, Ti,  and Fe for BW K/M  giants. They found their stars to be enriched in O  and in Ti, suggesting a rapid chemical enrichment of the Galactic bulge. \citet{ryde2010} determined abundance ratios for 11 bulge giants, and found enhanced  $\rm  [O/Fe]$, $\rm [Si/Fe]$, and $\rm [S/Fe]$ ratios with increasing metallicity, up to approximately $\rm [Fe/H] \sim -0.3$. However, all of these studies were based on ``single-slit'' spectroscopy, leading to a limited number of objects. The APOGEE  survey (\citealt{majewski2015}) is the first large-scale near-IR, high-resolution  ($\rm R \sim 22\,500$) survey, with its 300 fibre spectrograph able to assemble, for the first time, large samples of bulge stars with this information.  


Our aim in this paper is to compare stellar parameters and individual abundances, for $\alpha$- and iron-peak elements, of the APOGEE ASPCAP pipeline (DR13) with respect to previously found  literature values for stars in BW. We also compare the relative location of the bulge sequence in the [$\alpha$/Fe] vs. [Fe/H] plane as seen by the APOGEE and GES surveys. The study of Galactic stellar populations is entering  the era of big data. Surveys such as APOGEE and GES  are producing massive internally homogeneous databases of parameters and abundance ratios for samples of several thousands of  stars.  In this context, it is of great importance to properly characterize the eventual scale differences between results provided by different surveys for stars in common. 


\section{The samples}

\begin{figure*}[!ht]
  \centering
  \includegraphics[width=0.92\textwidth,angle=0]{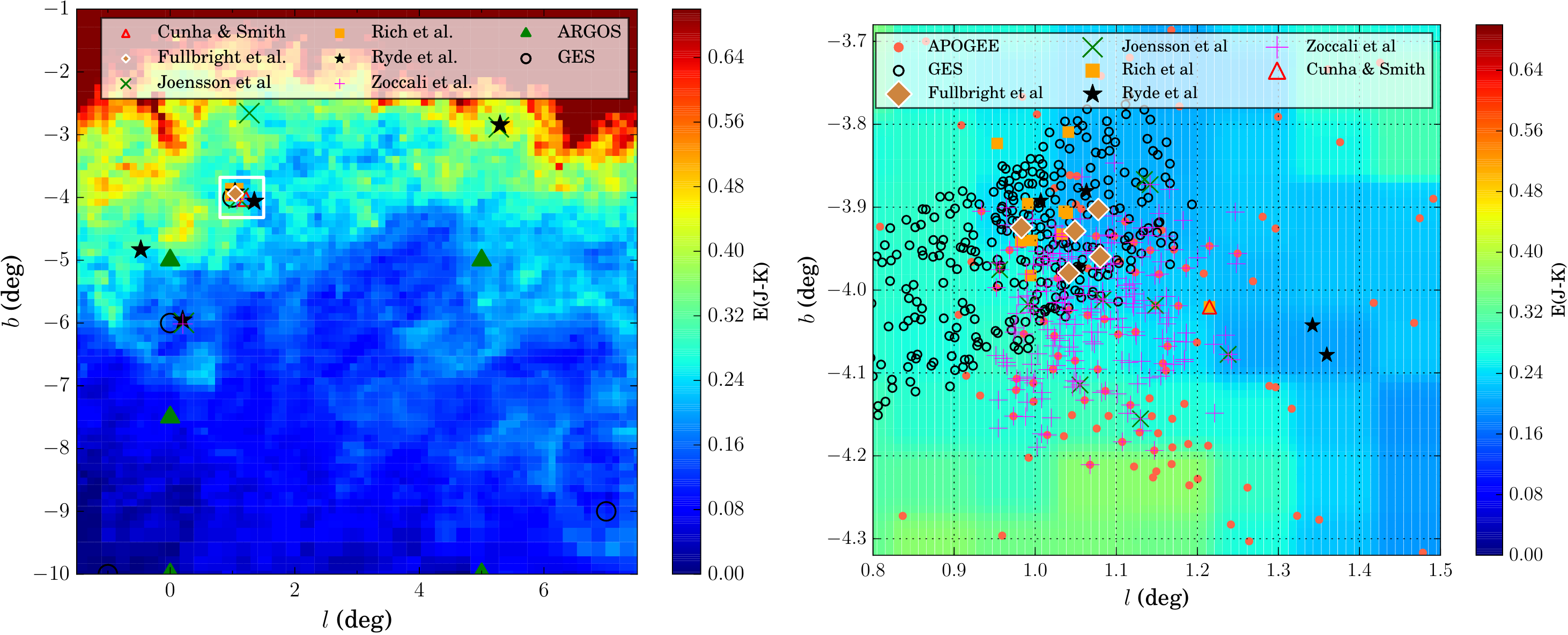}
  \caption{Baade's window finding chart. \textit{Left panel: }the white box highlights the position of Baade's window in a broader bulge area depicted by an extinction map, as obtained from the results of \citet{gonzalez2012}. The approximate position of other studied bulge fields, including those of the ARGOS and GES surveys, are depicted by several different symbols, as indicated in the figure. \textit{Right panel: }zoom of the highlighted area around Baade's window, showing the position of individual stars from the studies compared to APOGEE in this paper. Note that the distribution of APOGEE targets span a larger area, so that the targets displayed here are a subset of those adopted in this study.}
  \label{fig:finding_chart}
  \end{figure*}

\subsection{APOGEE} \label{APOGEE}
The Apache Point Observatory Galactic Evolution Experiment (APOGEE; \citealt{majewski2015}), began as one of four Sloan Digital Sky Survey-III (SDSS-III, \citealt{eisenstein2011}) experiments. It is a large-scale, near-IR,  high-resolution ($R \sim 22\,500$) spectroscopic survey of Milky Way stellar populations (\citealt{zasowski2013}). The survey uses a dedicated, 300-fibre, cryogenic spectrograph coupled to the wide-field  Sloan 2.5\,m telescope (\citealt{gunn2006}) at the Apache Point Observatory (APO).
APOGEE observes in the $H$-band, ($1.5\,\mu{\rm m}-1.7\,\mu{\rm m}$),  where extinction by dust is significantly lower than at optical wavelengths ($\rm A_{V}/A_{H} = 6$). Stellar parameters and chemical abundances for up to 22 elements can be determined by the Apogee Stellar Parameters and Chemical Abundances Pipeline (ASPCAP, \citealt{garcia16}). These values are based on a $\chi^{2}$ minimization between observed and synthetic model spectra (see \citealt{zamora2015}; \citealt{holtzman2015} for more details),  performed with the {\sc FERRE} code \citep[][and subsequent updates]{allende2006}. Several improvements in the data processing and analysis as presented in  Data Release 13 (DR13, \citealt{dr13}) have been accomplished: e.g., the variation of the line-spread function (LSF) as a function of the fiber number has been taken into account. A new updated line list has been used for determining stellar parameters and abundances, as well as a new MARCS model atmosphere grid going down to $\rm T_{eff} < 3500\,K$. In addition, separate synthetic spectral grids are used for dwarfs and giants, and updated relationships for the micro-turbulence and macro-turbulence have been adopted. For more details we refer the interested reader to Holtzman et al. (2017, in prep.) and the DR13 online documentation.

Baade's window has been observed in two Sloan fiber plug-plates (5817 and 5818), resulting in a total of 425 unique objects. A total of  287 stars were observed in the regular target selection mode of APOGEE (extratarget flag ==0) with the usual color cut of $\rm (J-K)_{0} > 0.5$ \citep[see][]{zasowski2013},  as well as 138 stars came from a commissioning plug-plate which have lower quality.  For stellar parameter estimates we used the   calibrated stellar parameters (PARAM) and individual chemical abundances ([X/Fe]), which were calibrated  using a sample of well-studied field and cluster stars, including a large number of stars with asteroseismic stellar parameters from NASA’s Kepler mission (see \citealt{holtzman2015} and Holtzman et al. 2017, in prep.). Using calibrated parameters implies a limit in $\rm log\,g < 3.8$.  The abundances have been calibrated internally, using data for stars  in open clusters in an attempt to remove abundance trends with effective temperature. In addition, the abundances have now been calibrated externally, in order to go through $\rm  [X/Fe]=0$ at $\rm  [Fe/H]=0$  for local thin disk stars.

 In order to retrieve the stellar parameters and the chemical abundances of the Baade's window field, one has to download the corresponding fits table file\footnote{\url{http://www.sdss.org/dr13/irspec/spectro_data/allStar-l30e.2.fits}} where all relevant information such as stellar parameters ($\rm T_{eff}$, log\,g, [Fe/H],$\rm [\alpha/Fe] $), chemical abundances, photometry, etc. are stored. The detailed description of this data file can be found on the corresponding webpage\footnote{\url {https://data.sdss.org/datamodel/files/APOGEE_REDUX/APRED_VERS/APSTAR_VERS/ASPCAP_VERS/RESULTS_VERS/allStar.html}}. Note that both uncalibrated and calibrated stellar parameters as well as chemical abundances
are given in this file. For BW, the easiest way to retrieve  the 425 objects is to search for the field name $\rm ``BAADEWIN"$
In the left panel of Fig.~\ref{fig:finding_chart}, the position of Baade's window is highlighted in a broader view of the bulge region. The background color map depicts the reddening variations according to the results of \citet{gonzalez2012}. The position of several fields studied in the literature, including the ARGOS and GES surveys, are depicted to put BW in the context of other bulge studies.

\begin{figure}[!htbp]
  \centering
        \includegraphics[width=0.49\textwidth,angle=0]{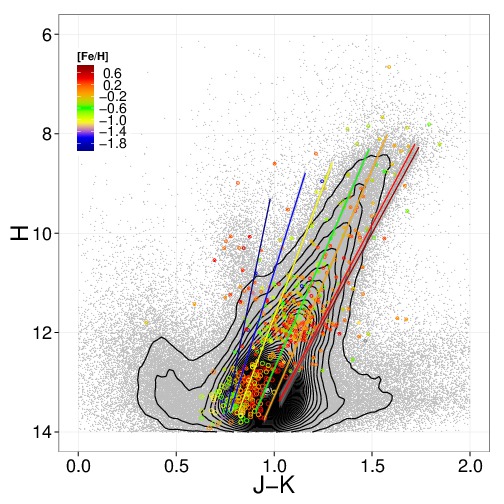}
        \caption{2MASS H vs. J--K diagram for stars in  BW. Superimposed over the  individual star measurements are the density contours. The APOGEE targets are shown as  filled circles; the GES targets are  shown as open circles. The colour scale indicates the metallicity of the stars. Superimposed are the 10\, Gyr PARSEC isochrones for  different metallicities, colour coded as given in the legend.}
  \label{CMD_FeH}
  \end{figure}

Figure~\ref{CMD_FeH} shows the 2MASS color-magnitude  diagram  (H vs. J--K) for Baade's window. The APOGEE and DR1-GES  samples are  colour coded as a function of metallicity. We also show the PARSEC isochrones (\citealt{bressan2012}), assuming an age of 10\,Gyr,  with varying metallicities assuming that the stars in BW have a mean distance of 8.4\,kpc (\citealt{GCdistance}). We assume an average extinction of $\rm A_{V}= 1.5$ (\citealt{stanek1996}), and the reddening law of \citet{nishiyama2009}, which we used to  redden the isochrones.  Stars with  $\rm log\,g > 3.5$ have been excluded.  We see clearly that the APOGEE stars have redder $\rm (J-K)_{0}$ colours than
predicted by the most metal-rich isochrones (Z=0.07). While  the metal-poor stars have generally bluer (J--K) colours,  stars with similar spectroscopically  determined metallicities show a large spread in the (J--K) colour. 

\subsection{The comparison samples}

As mentioned in Sect.~\ref{APOGEE}, BW was targeted by APOGEE for calibration  with respect to studies having well-established stellar parameters and chemical abundances. For the cross-identification of coordinates and 2MASS ID's of our stars, we  used Table\,1 of \citet{church2011},  who provided coordinates and 2MASS names for all stars in Arp's finding chart (\citealt{Arp1965}).  \citet{fulbright2006} obtained effective temperatures based  on V--K data, together with the photometric transformation to $\rm T_{eff}$ from \citet{alonso1999}.
We have in total  five stars in common with \citet{fulbright2006} (hereafter referred as Ful06), four stars with \citet{rich2005} (hereafter referred as Rich05), three with \citet{ryde2010} (hereafter referred as Ryde10), one with \citet{cunha06} (hereafter referred as Cunha06), and 55 with \citet{zoccali08} (hereafter referred as Zocc08). 
The right panel of Fig.~\ref{fig:finding_chart} displays the position of the individual stars in all these studies. The region displayed corresponds to that highlighted by a white square in the left panel of the same figure.
Table~\ref{table1} lists the stellar parameters of the different literature values, as well as those from APOGEE (DR13, \citealt{dr13}).
\begin{table*}[!htbp]
\caption{Stellar parameters comparison. Columns 2, 3, and 4 list the stellar parameters as derived in the reference quoted in the last column. The APOGEE stellar parameters are listed in columns 5, 6, and 7.} 
\label{table1}
\begin{tabular}{ccccccccc}
\hline \multicolumn{1}{|c|}{\textbf{2MASSid}} & \multicolumn{1}{c|}{\textbf{$\rm T_{eff}$}} & \multicolumn{1}{c|}{\textbf{$\rm log\,g$}} & \multicolumn{1}{c|}{\textbf{$\rm [Fe/H]$}} &\multicolumn{1}{c|}{{Colour}} & \multicolumn{1}{c|}{\textbf{$\rm T^{APOGEE}_{eff}$}} & \multicolumn{1}{c|}{\textbf{$\rm log\,g^{APOGEE}$}} & \multicolumn{1}{c|}{\textbf{$\rm [Fe/H]^{APOGEE}$}} & \multicolumn{1}{c|}{\textbf{Ref}} \\ \hline 

\hline
18033576-2958353&4257&1.55&-0.37&2.929\tablefootmark{a} &4179&1.79&-0.36&Ful06\\
18034989-3000083&4184&1.53&0.16&3.050\tablefootmark{a} &4251&2.42&0.28&Ful06\\
18033819-3000515&4097&0.87&-1.15&3.269\tablefootmark{a} &4411&2.07&-0.89&Ful06\\
18034938-3002440&4554&1.66&-1.22&2.552\tablefootmark{a} &4392&1.52&-1.17&Ful06\\
18032824-3004108&4157&1.58&-0.41&3.124\tablefootmark{a} &4090&1.78&-0.38&Ful06\\
18032356-3001588&3902&0.51&-1.25&3.776\tablefootmark{b}&3994&1.40&-1.04&Ryde10\\
18032843-2958416&4197&1.29&-0.90&3.103\tablefootmark{b}&4202&1.67&-0.85&Ryde10\\
18034939-3001541&4106&0.89&-0.23&3.131\tablefootmark{b}&4194&1.86&-0.08&Ryde10\\
18042265-2954518&3200&0.50&-0.15&1.370\tablefootmark{b}&2994&-0.06&-0.16&Rich05\\
18042265-2954518&3375&0.40&-0.05&1.370\tablefootmark{b}&2994&-0.06&-0.16&Cunha06\\
18030810-2957480&3200&0.50&-0.17&1.300\tablefootmark{b}&3080&-0.23&-0.49&Rich05\\
18033169-3000437&4000&1.00&-0.09&0.840\tablefootmark{b}&3767&1.24&-0.21&Rich05\\
18033374-3003572&4000&1.00&-0.03&0.840\tablefootmark{b}&3853&1.25&-0.29&Rich05\\
18034320-2959404&4850&1.93&-0.37&1.667\tablefootmark{c}&4298&1.78&-0.76&Zocc08\\
18034616-2958303&4000&1.52& 0.51 & 2.083\tablefootmark{c} &4124& 2.21&0.38&Zocc08\\
18040747-2955001&4250&1.70& 0.05 & 2.050\tablefootmark{c} &4128& 2.10&0.20&Zocc08\\
18040026-2958253&4850&2.00&-0.67 &1.690\tablefootmark{c}&4398&1.87&-0.94&Zocc08\\
18041720-2956497&4400&1.84&-0.05 & 2.081\tablefootmark{c}&4156&1.83&-0.18&Zocc08\\
18043406-2959546&4930&2.04&-0.10 &1.636\tablefootmark{c}&4207&2.43&0.11&Zocc08\\
18042724-3001108&5100&2.06&-0.22&1.527\tablefootmark{c}&3270&-0.37& -2.15&Zocc08\\
18042722-2958570&4700&1.94&-0.20&1.793\tablefootmark{c}&4327&2.19 &-0.42&Zocc08\\
18040883-3002037&4200&1.67&-0.24&2.036\tablefootmark{c}&4092&1.83&-0.26&Zocc08\\
18041554-3001431&4550&1.81&-0.45&1.814\tablefootmark{c}&4210&1.80&-0.63&Zocc08\\
18041486-3010159&4850&2.12&0.45 &1.700\tablefootmark{c}&3810&1.86&0.34&Zocc08\\
18041410-3007315&4550&1.87&-0.04&1.925\tablefootmark{c}&4193&2.15&-0.25&Zocc08\\
18035937-3006027&4500&1.91&-0.60&1.931\tablefootmark{c}&4172&2.09&-0.64&Zocc08\\
18033691-3007047&4600&1.97&0.05&2.017\tablefootmark{c}&4157&1.90&-0.21&Zocc08\\
18041187-3006214&4300&1.74&-0.22&2.072\tablefootmark{c}&4109&1.83&-0.29&Zocc08\\
18041606-3005254&4200&1.67&0.40 &2.029\tablefootmark{c}&4229&2.28 &0.37 &Zocc08\\
18033660-3002164&4950&2.00&-1.05&1.664\tablefootmark{c}&4927&2.23&-1.12&Zocc08\\
18034092-3004423&4650&1.99&-0.21&1.858\tablefootmark{c}&4313&2.78&-0.46&Zocc08\\
18034906-3003384&4500&1.87&-0.32&1.876\tablefootmark{c}&4252&1.83&-0.49&Zocc08\\
18035632-2956410&4350&1.84&0.27&2.255\tablefootmark{c}&4173&2.42&0.53&Zocc08\\
18035131-2957281&4400&1.91&-0.08&2.192\tablefootmark{c}&4075&1.96&-0.14&Zocc08\\
18034379-2957162&4350&2.03&0.22&2.296\tablefootmark{c}&3970&2.09&-0.36&Zocc08\\
18041178-2951100&4400&1.86&0.15&2.028\tablefootmark{c}&4237&2.33&0.22&Zocc08\\
18040492-2952427&4400&1.91&0.29&2.133\tablefootmark{c}&4232&2.48&0.20&Zocc08\\
18040398-2959223&4100&1.70&-0.15&2.255\tablefootmark{c}&3857&1.09&-0.48&Zocc08\\
18035929-2949519&4150&1.87&0.28&2.520\tablefootmark{c}&4011 &1.78&0.33&Zocc08\\
18040008-2955079&4500&1.96&-0.05&2.077\tablefootmark{c}&4165&2.17&-0.26&Zocc08\\
18041328-2958182&4300&1.87&0.25&2.192\tablefootmark{c}&4045 &1.94&0.24 &Zocc08\\
18043142-2959515&4200&1.76&0.17&2.338\tablefootmark{c}&3854 &1.36&-0.49&Zocc08\\
18041770-3000304&4150&1.77&0.28&2.286\tablefootmark{c}&4087 &2.27&0.31 &Zocc08\\
18045455-2958169&4550&1.95&-0.19&2.017\tablefootmark{c}&4173&1.69&-0.22&Zocc08\\
18042994-3004324&4400&1.86&-0.11&2.074\tablefootmark{c}&4144&2.00&-0.21&Zocc08\\
18045547-3003285&4300&1.98&0.43 &2.435\tablefootmark{c}&4039 &1.78&0.14&Zocc08\\
18044764-3005147&4600&2.04&-0.65&2.234\tablefootmark{c}&4149&1.66&-0.78&Zocc08\\
18043821-3003251&4350&1.99&0.34&2.452\tablefootmark{c}&4163&2.46 & 0.35&Zocc08\\
18042236-3004162&4400&1.98&0.49&2.193\tablefootmark{c}&4168&2.37 & -0.02&Zocc08\\
18042178-3006128&4500&1.99&-0.25&1.959\tablefootmark{c}&4124&1.57&-0.50&Zocc08\\
18043319-3009500&4500&1.94&0.02&2.155\tablefootmark{c}&4210&1.96&-0.08&Zocc08\\
18042920-3006120&4150&1.76&0.38&2.256\tablefootmark{c}&4151&2.41&0.37&Zocc08\\
18044899-3008077&4200&1.87&0.14&2.511\tablefootmark{c}&4066&2.05&0.29&Zocc08\\
18041165-3009495&4400&1.91&0.46&2.193\tablefootmark{c}&4049&1.87&0.30&Zocc08\\
18040840-3004382&4250&2.00&0.12&2.544\tablefootmark{c}&3873&1.45&-0.036&Zocc08\\
18041566-3008540&4050&1.67&0.35&2.156\tablefootmark{c}&4176&2.30&0.36&Zocc08\\
18044899-3008077&4200&1.87&0.14&2.511\tablefootmark{c}&4066&2.05&0.29&Zocc08\\
18041165-3009495&4400&1.91&0.46&2.193\tablefootmark{c}&4049&1.87&0.30&Zocc08\\
\hline
\end{tabular}
\end{table*}
\begin{table*}[!htbp]
\begin{tabular}{ccccccccc}
\multicolumn{8}{c}%
{{\bfseries \tablename\ \thetable{} -- continued from previous page}} \\
\hline \multicolumn{1}{|c|}{\textbf{2MASSid}} & \multicolumn{1}{c|}{\textbf{$\rm T_{eff}$}} & \multicolumn{1}{c|}{\textbf{$\rm log\,g$}} & \multicolumn{1}{c|}{\textbf{$\rm [Fe/H]$}} &\multicolumn{1}{c|}{{Colour}} &  \multicolumn{1}{c|}{\textbf{$\rm T^{APOGEE}_{eff}$}} & \multicolumn{1}{c|}{\textbf{$\rm log\,g^{APOGEE}$}} & \multicolumn{1}{c|}{\textbf{$\rm [Fe/H]^{APOGEE}$}} & \multicolumn{1}{c|}{\textbf{Ref}} \\ \hline 
18040840-3004382&4250&2.00&0.12&2.544\tablefootmark{c}&3873&1.45&-0.036&Zocc08\\
18041566-3008540&4050&1.67&0.35&2.156\tablefootmark{c}&4176&2.30&0.36&Zocc08\\
18042205-3011214&4350&1.89&0.09&2.113\tablefootmark{c}&4159&2.04&0.20&Zocc08\\
18041492-3007088&4400&2.01&-0.01&2.231\tablefootmark{c}&3893&1.421&-0.20&Zocc08\\
18035975-3007473&4350&1.88&0.31&2.256\tablefootmark{c}&4046 &1.84&0.24&Zocc08\\
18040535-3005529&4100&1.84&0.35&2.530\tablefootmark{c}&4049 & 2.27&0.50&Zocc08\\
18034164-3007549&4200&1.80&-0.12&2.407\tablefootmark{c}&3959&1.805&0.19&Zocc08\\
18040486-3003009&4500&2.00&-0.20&2.032\tablefootmark{c}&4105&1.87&-0.61&Zocc08\\
18040030-3004017&4300&1.85&0.08 &2.157\tablefootmark{c}&3964 &1.89&-0.73&Zocc08\\
18031997-3004274&4300&1.98&-0.17&2.398\tablefootmark{c}&3849&1.26&-0.30&Zocc08\\
18034052-3003281&4450&1.90&0.55 &1.857\tablefootmark{c}&4486 &2.24&-0.90&Zocc08\\
18031683-3006111&4650&2.02&-0.05&1.959\tablefootmark{c}&4181 &1.95&-0.34&Zocc08\\
18033286-3005450&4200&1.94& 0.28 &2.384\tablefootmark{c}&4037  &2.11&0.34&Zocc08\\
18040170-3001491&4600 &1.99&-0.01&1.954\tablefootmark{c}&4118 &2.65&-0.03&Zocc08\\
\hline
\end{tabular}
\tablefoot{
\tablefoottext{a}{$\rm (V-K)_{0}$ colour}
\tablefoottext{b}{$\rm (J-K)_{0}$ colour}
\tablefoottext{c}{$\rm (V-I)_{0}$ colour}
}
\end{table*}




\section{Distances}

The spectrophotometric distances and reddenings for the full sample were calculated by using the stellar parameters $\rm T_{eff}$, $\rm log\,g$, and  [Fe/H], together with 2MASS J, H, and $\rm K_{S}$ photometry and associated errors, to  simultaneously compute the most likely line-of-sight distance and reddening by isochrone fitting with a set of PARSEC isochrones. We use the same method as presented in Rojas-Arriagada et al. (2016), considering a set of isochrones in the age range from 1 to 13\, Gyr in 1\,Gyr steps and metallicities from -2.2 to +0.5  in steps of 0.1\,dex. We have chosen this method to calculate the  distances consistently with respect to the bulge sample of the GES. A comparison has been done with the distance code of the ``BPG group'' (\citealt{santiago2016}), which uses a Bayesian approach. We did not find any systematic offset in the distance distribution.

\begin{figure}[!htbp]
 \centering
 \includegraphics[width=0.48\textwidth,angle=0]{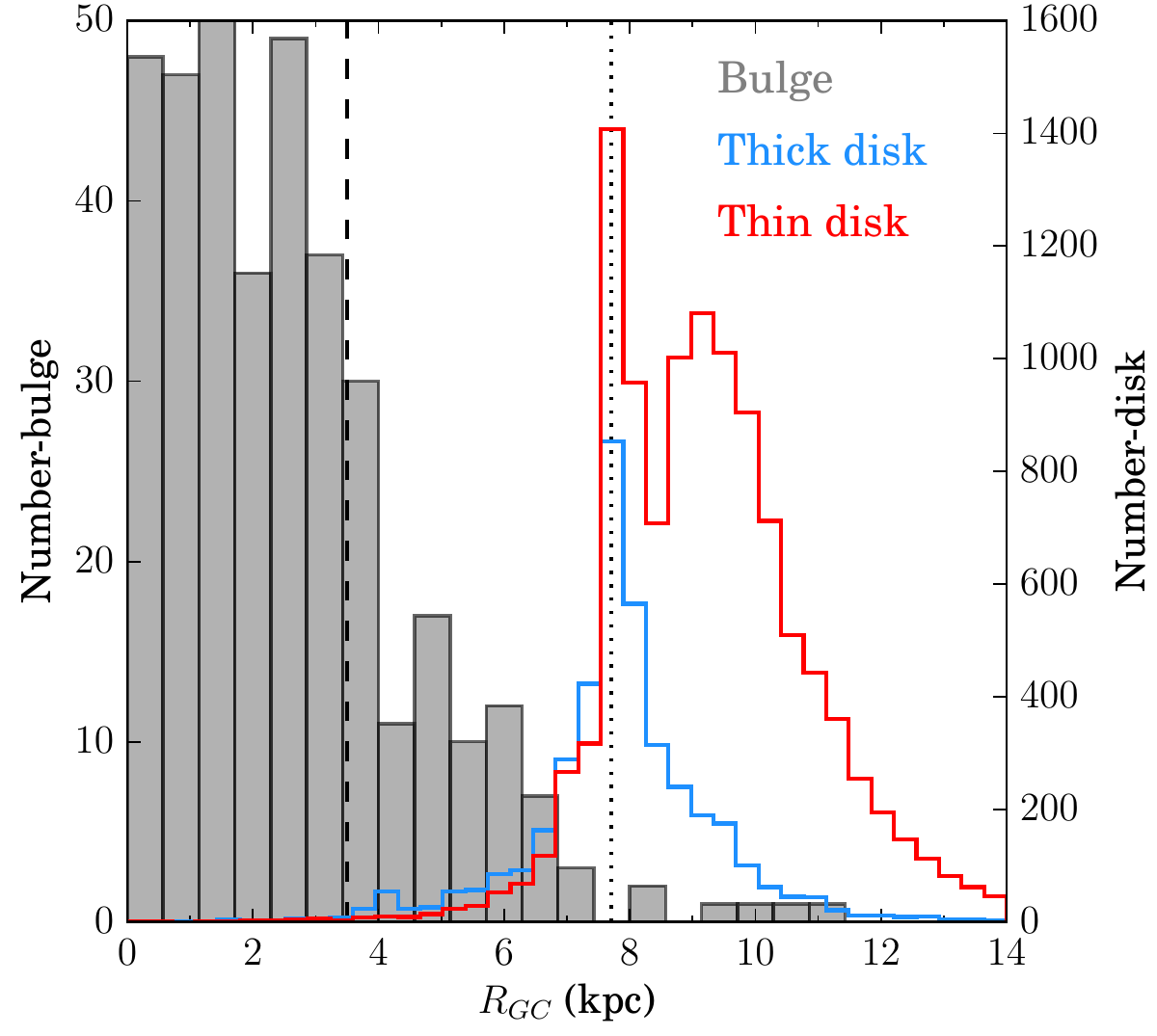}
 \caption{Distribution of the Galactocentric distances of our APOGEE sample. Grey bars indicate bulge stars (according to the left vertical scale), while the blue and red open histograms show the thick and the thin disc (according to the right vertical scale), respectively. The vertical dashed line defines our cut at $\rm R_{GC} = 3.5\,kpc$ to select likely bulge stars, while the vertical dotted line represents our cut at $\rm R_{GC} = 7.7\,kpc$ used to select the comparison disc sample (see Sect.~6).}
 \label{hist_distances}
 \end{figure}

Figure \ref{hist_distances} displays the histogram of our BW sample as gray bars.  The peak of the histogram is located close to the  distance of  the Galactic Center ($\rm R_{GC} =0\,kpc$), and then decreases toward larger distances.  For the discussion of the MDF (Sect.~\ref{MDF})  we restrict our sample to $\rm R_{GC} < 3.5\,kpc$, to ensure  having a  sample of stars likely located in the Galactic bulge.  This selection criterion leave us with 269 stars. The two open histograms  in Fig.~\ref{hist_distances} show the distribution of APOGEE disc stars, from which we selected a comparison sub-sample. The definition of this disc sample is discussed in Sect.~\ref{sec:abundancetrends}.

\section{Comparison of stellar parameters}

\begin{figure}[!htbp]
  \centering
        \includegraphics[width=0.40\textwidth,angle=0]{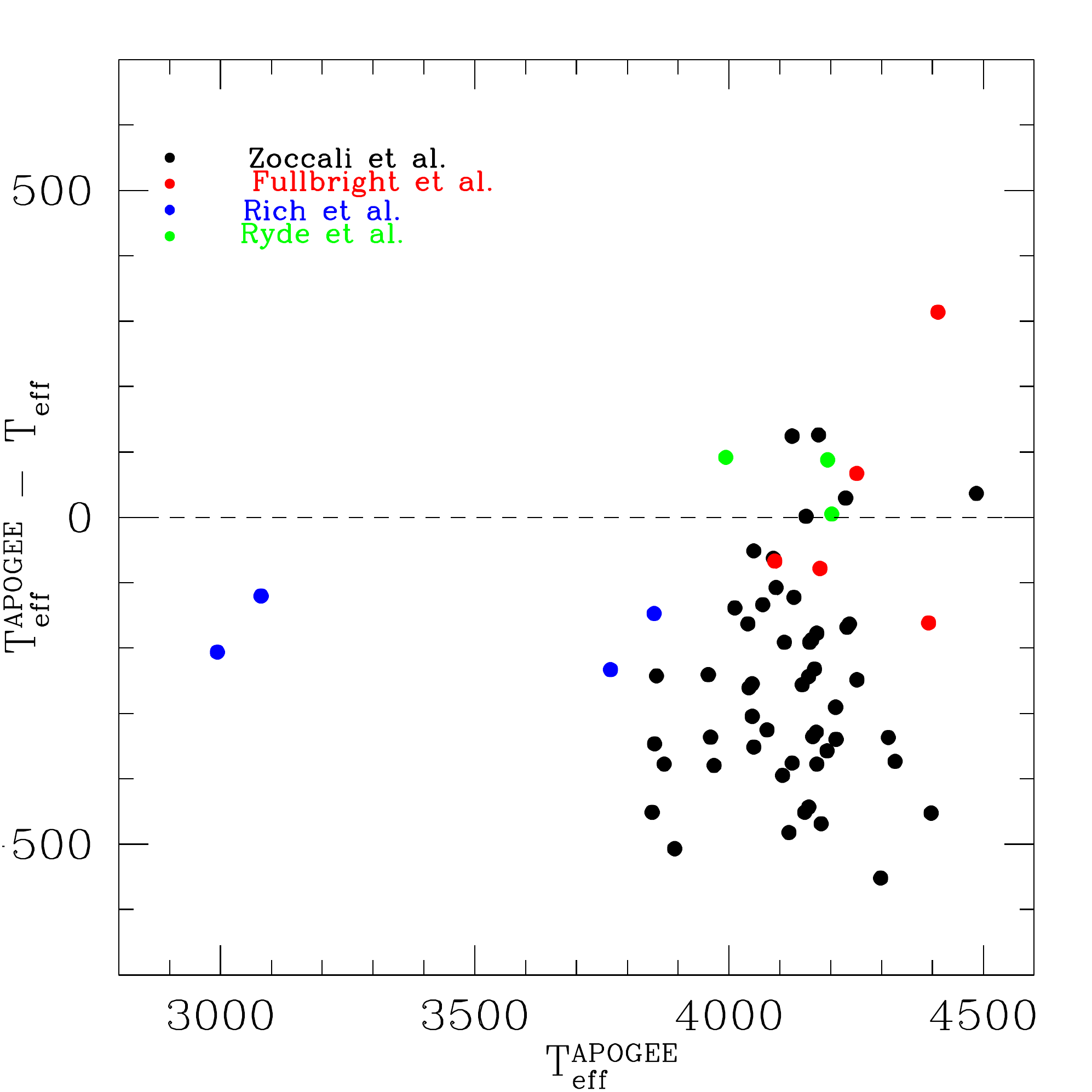}
        \caption{Difference in $\rm T_{eff}$  between APOGEE and literature values vs. $\rm T_{eff}$. Black circles  are giants from  \citet{zoccali08}, red circles are from \citet{fulbright2006}, blue circles are from \citet{rich2005}, and green circles are  from \citet{ryde2010}.}
  \label{DeltaTeff}
  \end{figure}

Figure~\ref{DeltaTeff} presents the comparison of the effective temperatures between Rich05, Ful06, Zocc08,  and Ryde10.
Table~2 gives the corresponding mean differences and the standard deviation of the stellar parameters with respect to APOGEE. The Rich05 and the Zocc08 stars exhibit systematically higher temperatures (177\,K and 255\,K, respectively) compared to APOGEE, while Ful06 and Ryde10  obtain  similar  temperatures (15\,K and 61\,K).  The largest dispersion appears for the Ful06 stars as well as for the  Zocc08 stars (186\,K and 161\,K, see Tab.~2),  while the dispersion in Ryde10 and Rich05 is rather small (49\,K and 52\,K). In our comparison sample (see Fig.~\ref{DeltaTeff}),  the effective temperatures of Ful06 were determined based on photometric  V--K colours, as well as from differential excitation temperatures and ionization temperatures. They found in general a very good agreement between these three temperature estimates with a small scatter.  Zocc08 uses V--I colours as a first estimate while the final Zocc08 values, which we adopted here for the comparison, were estimated spectroscopically by imposing excitation equilibrium on a set of $\sim60$ FeI lines. Rich05  estimated  temperatures based on J--K colours,  while Ryde10 uses the effective temperatures of Ful06.

One part of the large spread  in the photometric temperatures compared to the spectroscopic ones of APOGEE  can be explained by the inhomogenous use of photometric colours in the comparison work while the spectroscopic temperatures from APOGEE were determined in a homogeneous way.  We indicate in Tab.~1 the corresponding photometric colours as well as the photometric system used.  \citet{gonzalez2009} studied in detail the effective temperature scale using the infrared flux methode.  They show in their Tab.~5 that the V--K colours show the smallest dispersion in the determination of the effective temperature  ($\rm \sim 30\,K$) and should be therefore used for photometric temperaturs.  The  V--I colours  and especially the J--K  colours  show on the other hand  a much larger dispersion in the effective temperature. Clearly much more effort is needed to understand the difference between spectroscopic and photometric temperature determinations.

\begin{figure}[!htbp]
  \centering
         \includegraphics[width=0.40\textwidth,angle=0]{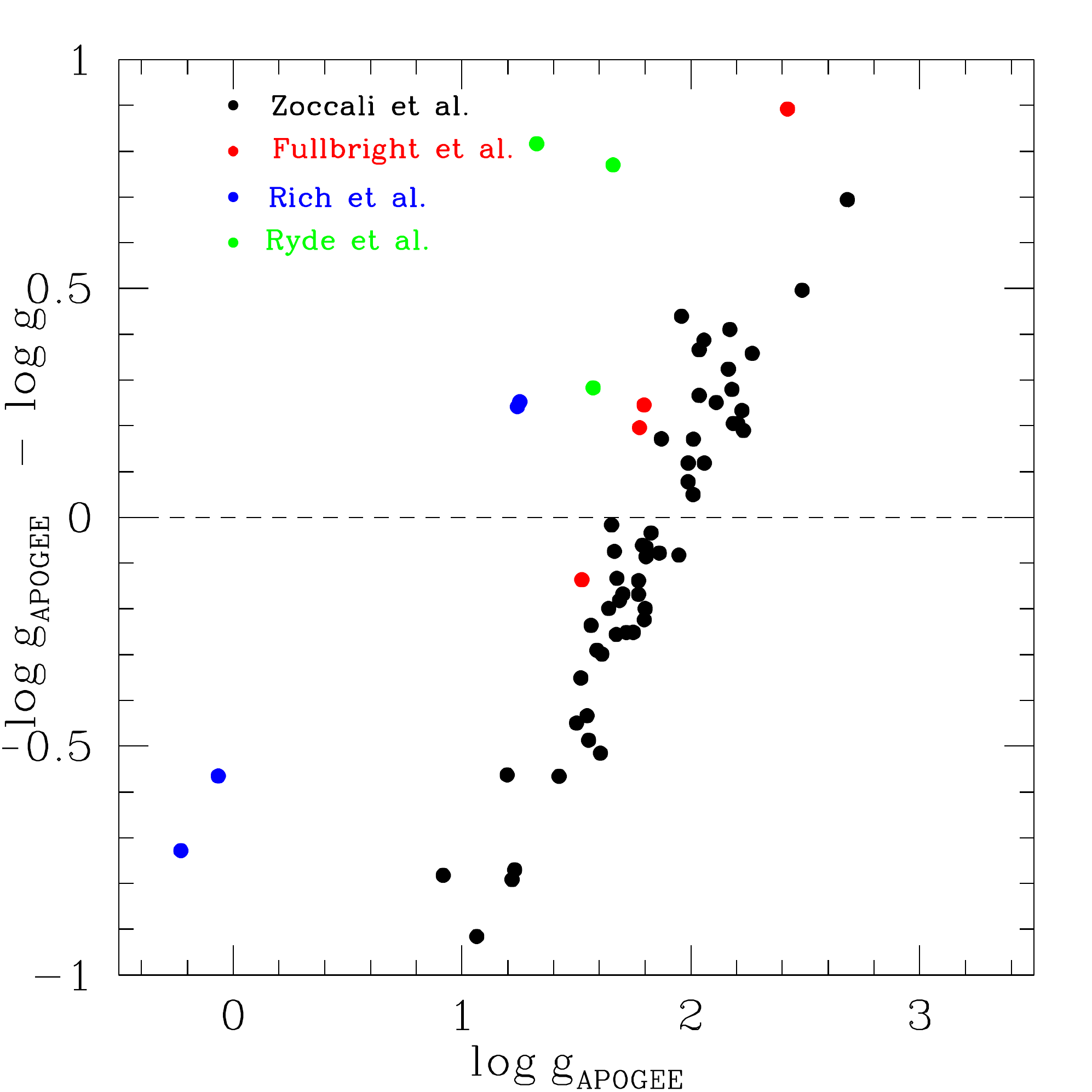}
        \caption{Difference in $\rm log\,g$  between  APOGEE and literature values  vs. $\rm log\,g$. Same symbols as in Fig.~\ref{DeltaTeff}.  A clear linear relation is seen for the Zocc08  sample  between the  difference of the spectroscopic and photometric gravities with respect to the spectroscopic log\,g values  (see text).}
  \label{Deltalogg}
  \end{figure}

Figure~\ref{Deltalogg} shows the APOGEE-literature comparison of surface gravity. The surface gravities of APOGEE are calibrated with respect to the asteroseismic log\,g estimates  from NASA's {\it{Kepler}} mission (\citealt{borucki2010}).  The  applied offset is about $\rm 0.2\,dex$ for stars with
 solar metallicities.  The typical dispersion between photometric log\,g and spectroscopic estimates are on the order of $\rm \pm 0.5\,dex$, while the dispersion in Ryde10 is slightly lower (0.3\,dex) although they exhibit a large offset (0.62\,dex).  In general, there is  a large systematic discrepancy between photometrically derived log\,g and spectroscopic estimates.

\begin{figure}[!htbp]
  \centering
         \includegraphics[width=0.48\textwidth,angle=0]{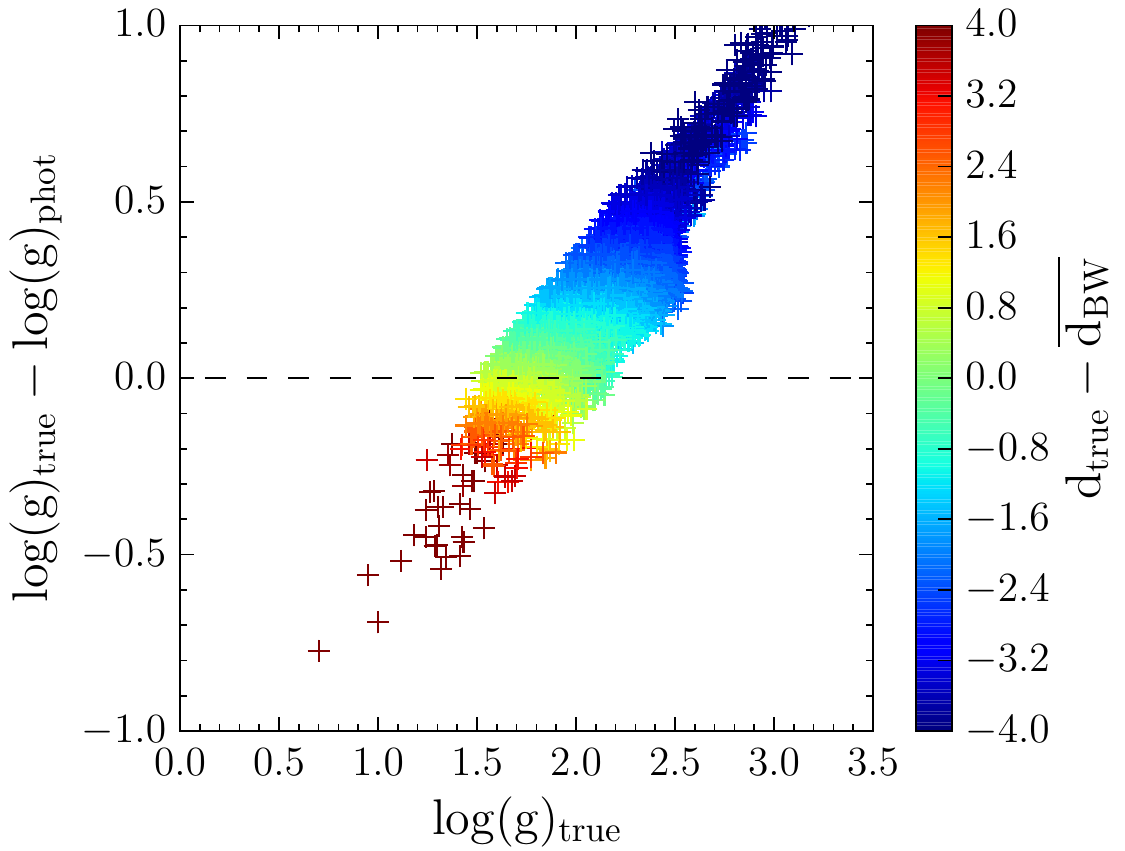}
        \caption{Difference in $\rm log\,g$ for stars selected from a TRILEGAL simulation of BW as a function of log\,g (true). The photometric values were computed by assuming all of the sample is at a distance of 8~kpc. Symbols are colour coded according to the difference of the true distance with respect to that assumption. Photometric gravities become smaller with respect to the true spectroscopic surface gravities if the assumed distance is shorter than 8\,kpc.}
  \label{Deltalogg_trilegal}
  \end{figure}

The Zocc08 photometric log\,g estimates  display a clear linear behavior  with respect to APOGEE. In Zocc08, the photometric gravities were estimated by assuming a mean stellar mass of $M=0.8~M_\odot$ and a sample distance of 8~kpc. The dispersion of their logg is about 0.25\,dex due to the intrinsic depth of the galactic bulge (\citealt{lecureur07}).  We investigate the effects of these assumptions by selecting stars from a TRILEGAL \citep{girardi2012} simulation using the Zocc08 photometric selection function (i.e., their Fig.~1), and calculating photometric log\,g values. Figure~\ref{Deltalogg_trilegal} displays the selected simulated stars in the same plane as in Fig.~\ref{Deltalogg}, colour coded by the difference of their true distances with respect to an assumed mean field distance ($\overline{d_{BW}}$). Stars at larger distances than $\overline{d_{BW}}$ are intrinsically luminous (low log\,g values) but appear fainter. By assuming these stars are at $\overline{d_{BW}}$, the resulting photometric log\,g values are larger, to account for their apparent low luminosity, determining a $\Delta log\,g (true-phot)<0$. Conversely, foreground stars at distances shorter than $\overline{d_{BW}}$, are on average intrinsically less luminous (high log\,g values). By assuming these stars are at $\overline{d_{BW}}$, the resulting  photometric log\,g values are smaller, to account for their apparent high luminosity, determining a $\Delta log\,g (phot-spec)>0$. The interplay between these complementary effects determine the linear trend observed in Fig.~\ref{Deltalogg} and reproduced in Fig.~\ref{Deltalogg_trilegal}. The dispersion around the mean comes from the fact that for a given distance interval there are stars spanning a range of true log\,g values. If we consider that the reddening in BW is small and homogeneous, photometric log\,g estimates in other windows of the Galactic bulge, where interstellar extinction is higher and spatially patchy, could  lead to even larger uncertainties. 

\begin{figure}[!htbp]
  \centering
        \includegraphics[width=0.40\textwidth,angle=0]{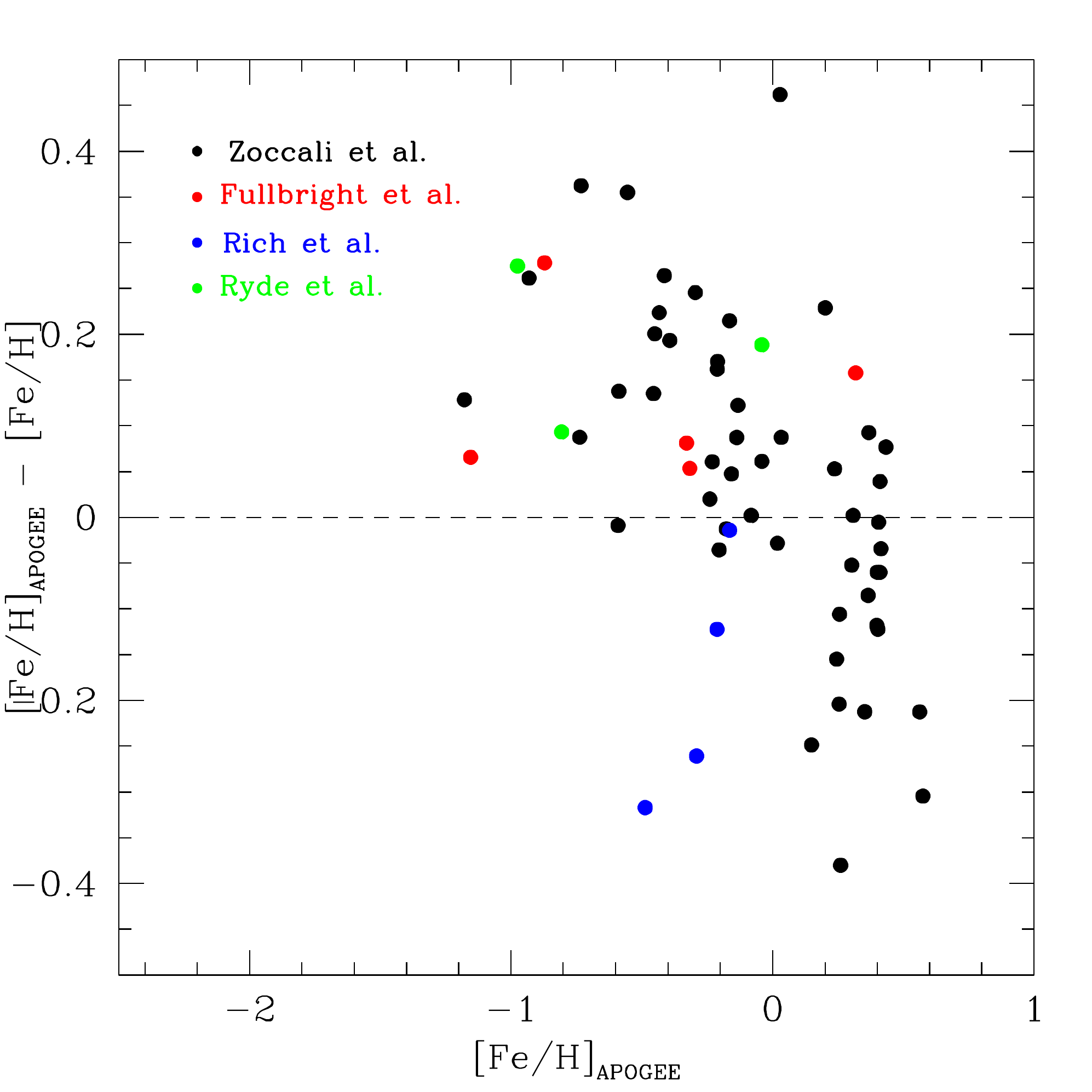}
        \caption{Difference in $\rm [Fe/H]$  between APOGEE and literature values vs. $\rm [Fe/H]$. Same symbols as in Fig.~\ref{DeltaTeff}. The typical dispersion is about 0.1\,dex,  while
the dispersion increases significantly for the Zocc08 sample ($\rm \sim 0.28\,dex$).}
  \label{DeltaFeH}
  \end{figure}

Figure~\ref{DeltaFeH} shows the comparison of the global metallicities. Except for Rich05, the comparison samples predict generally  too low metallicities with respect to APOGEE. The offset can go up to 0.18\,dex (Ryde10). The typical dispersion is about 0.15\,dex,  but  is significantly larger for the Zocc08 sample (0.28\,dex), where stars with $\rm [M/H] < 0$ are systematically more  metal-poor with respect to the APOGEE measurements. This larger scatter is partially due to the larger dispersion in $\rm T_{eff}$ and $\rm log\,g$ (see Fig.~\ref{DeltaTeff} and Fig.~\ref{Deltalogg}) of the Zocc08 determinations.

For the comparison of the $\alpha$-elements, we used the combination of Mg and Si for  Rich05 and Ful06, the \citet{gonzalez11a} values for the Zocc08 sample, and the global $\alpha$-element estimate for Ryde10. In general, the $\alpha$-element abundances in APOGEE are  lower compared to the literature values with differences ranging from 0.07\,dex (Zocc08) up to 0.19\,dex (Rich05, Ful06). The dispersion is about 0.1\,dex, where again Zocc08  has a larger dispersion (0.15\,dex).

\begin{figure}[!htbp]
 \centering
   \includegraphics[width=0.49\textwidth,angle=0]{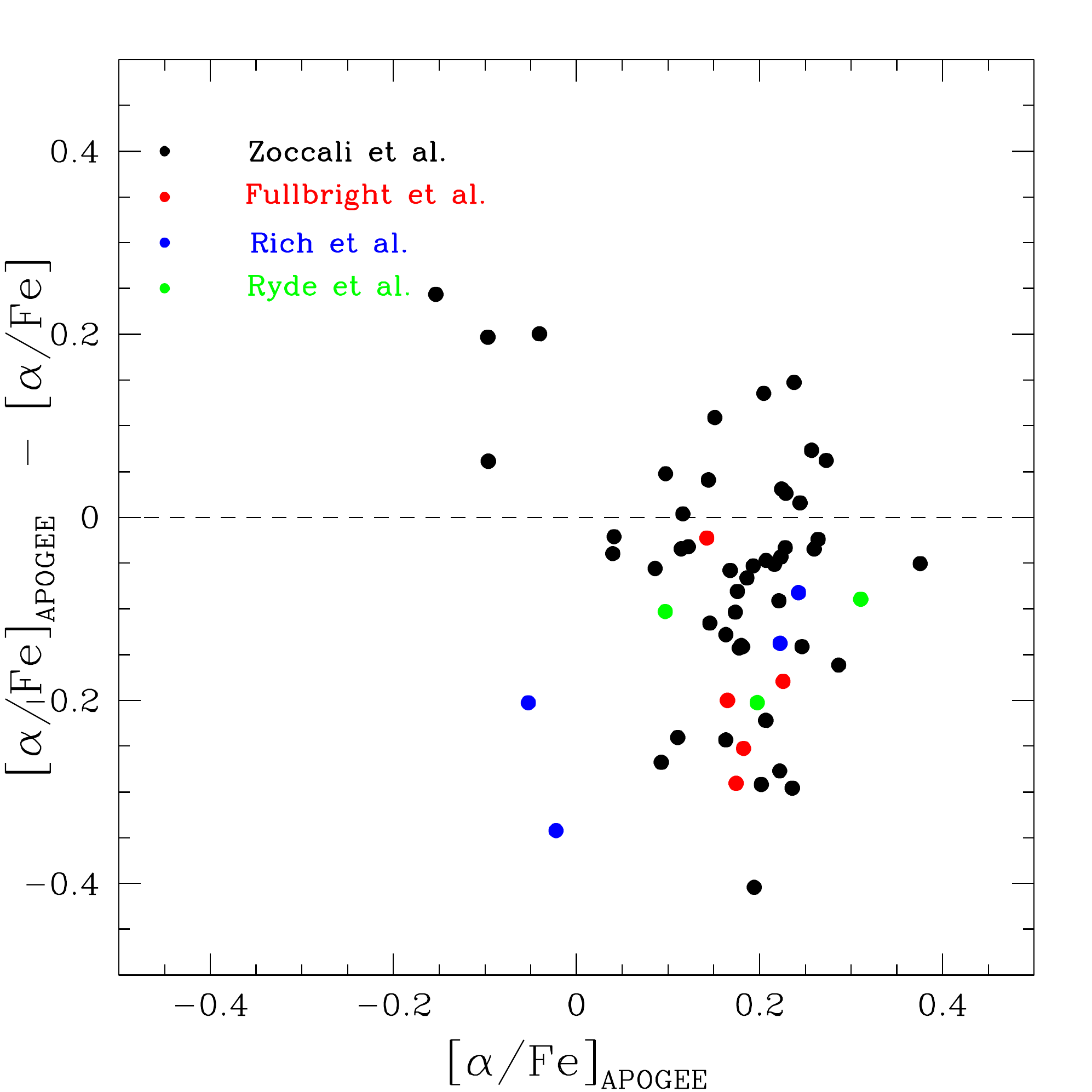}
   \caption{Difference in $\rm [\alpha/Fe]$  between APOGEE and  literature values  vs. $\rm [\alpha/Fe]$. The symbols are the same  as in Fig.~\ref{DeltaTeff}. The typical dispersion is $\rm 0.1\,dex$, with a slightly higher value for the Zocc08 sample ($\rm \sim 0.15\,dex$). }
 \label{Deltaalpha}
 \end{figure}

\begin{table}
\begin{tabular}{cccccc}
                           &  Rich05 &  Ful06  & Zocc08  & Ryde10  &   \\
\hline
$\rm <\Delta T_{eff}>$     &   -177  &     15  &   -255  &     61  &   [K]\\
$\rm \sigma (T_{eff})$     &     52  &    186  &    161  &     49  &   [K] \\
$\rm <\Delta log\,g>$      &   -0.19  &   0.48  &  -0.12  &  0.62  &   [dex]\\
$\rm \sigma (log\,g)$      &   0.52  &   0.54  &   0.48  &   0.29  &   [dex]\\
$\rm <\Delta [Fe/H]>$      &  -0.18  &   0.13  &   0.10  &   0.18  &   [dex]\\
$\rm \sigma ([Fe/H])$      &   0.14  &   0.09  &   0.28  &   0.09  &   [dex]\\
$\rm <\Delta [\alpha/Fe]>$ &  -0.19  &  -0.19  &  -0.07  &  -0.13  &   [dex]\\
$\rm \sigma ([\alpha/Fe])$ &   0.11  &   0.10  &   0.15  &   0.06  &   [dex]\\
\end{tabular}
\label{stat}
\caption{Difference between stellar parameters from APOGEE compared to the literature ($\rm \Delta$) and its  r.m.s. dispersions of the differences  ($\rm \sigma$) for stellar parameters $\rm T_{eff}$, $\rm log\,g$, $\rm [Fe/H]$, and $\rm \alpha$.}
\end{table}

\begin{figure}[!htbp]
{\center \includegraphics[width=0.49\textwidth,angle=0]{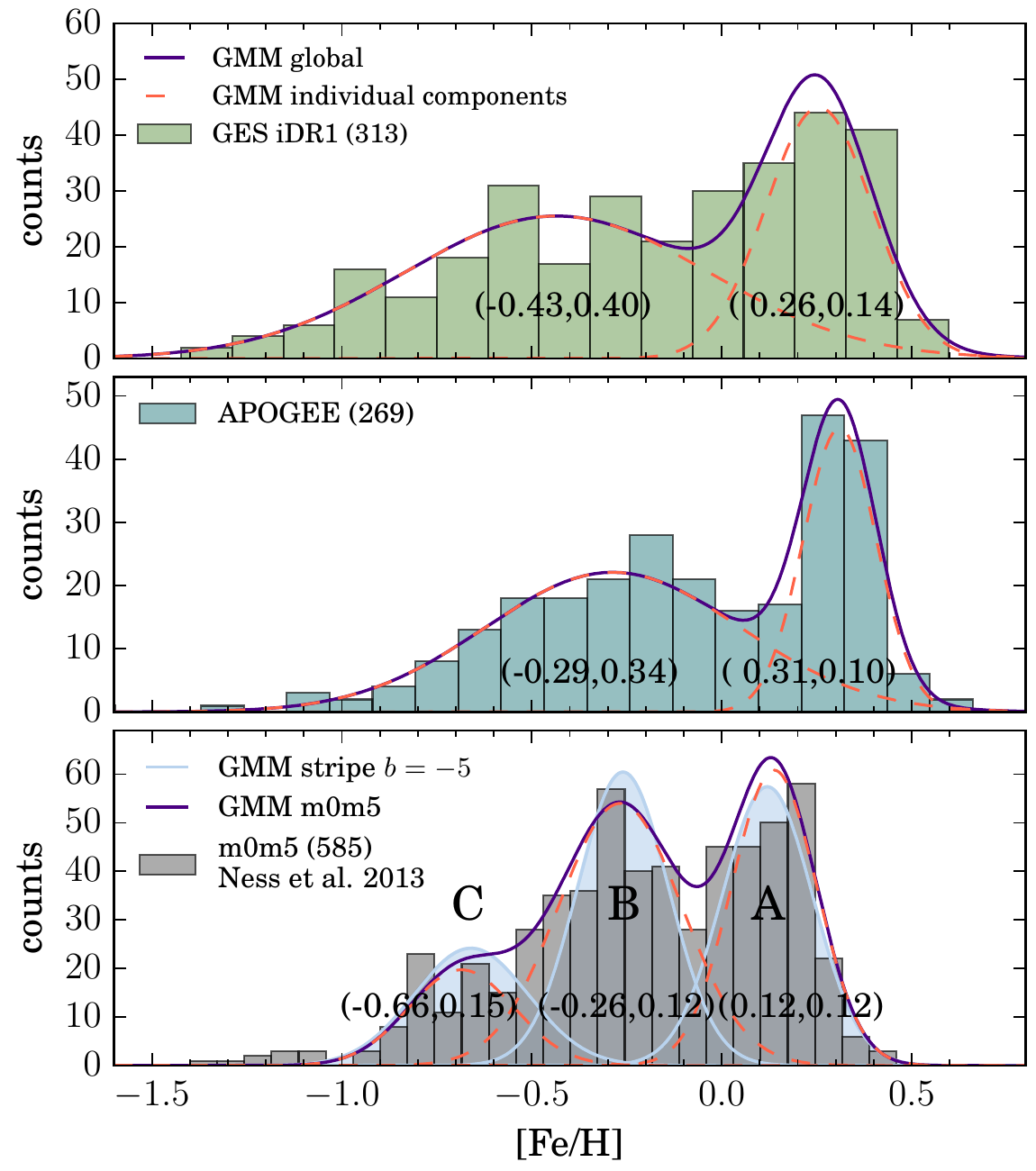}}
\caption{Metallicity distribution function for stars in  BW. Only sources with $\rm R_{GC} < 3.5\,kpc$ are included. \textit{Top panel: }MDF of the GES BW data (iDR1). \textit{Middle panel:}MDF of the APOGEE data. In each case, the GMM decomposition is depicted by dashed (individual components) and solid (global profile) lines, with mean and width values in parenthesis. \textit{Lower panel:} MDF of the closest field to BW from ARGOS data \citep{ness2013}. The GMM decomposition is depicted as in the other panels. For comparison, the GMM decomposition of  the $b=-5^\circ$ strip, as taken from \citet{ness2013}, is depicted by the three Gaussian profiles, with the corresponding mean and width values quoted in parenthesis.}
\label{fig:MDF}
\end{figure}

\section{The metallicity distribution function} \label{MDF}
The   MDF of BW stars  is an important tool to unravel the mix of stellar populations that comprise the Galactic bulge. With larger sample sizes, it becomes clear that
the MDF does not reflect a single stellar population, but reveals at least a bimodal nature. This behaviour was previously  noticed  by \citet{hill2011}, and further characterized by \citet{babusiaux2010}, who found metal-rich stars displaying bar-like kinematics in contrast with the isotropically hotter metal-poor bulge.

Figure~\ref{fig:MDF} compares the BW MDF as sampled by APOGEE with two other recent large-scale spectroscopic surveys, the ARGOS \citep[in m0m5, a field close to BW;][]{freeman2013,ness2013} and the GES  \citep[iDR1][]{gilmore2010,rojas2014,mikolaitis2014}. In each case, we used spectrophotometric distances to select samples of likely bulge stars as those with Galactocentric distance satisfying $\rm R_{GC} < 3.5\,kpc$, as done for the APOGEE sample (Sect.~3). The Visual inspection of the metallicity distributions of these likely bulge stars, as depicted by the histograms in each panel, reveals some qualitative differences. While the GES and APOGEE samples can be described with bimodal distributions , the ARGOS sample requires  apparently  a third component.

To quantify the substructure of the different MDFs, we perform a Gaussian mixture model (GMM) decomposition independently for each case. A GMM is a parametric probability density function given by a weighted sum of a number of Gaussian components. The GMM parameters are estimated as those that provide the best representation of the dataset density distribution structure. The Expectation-Maximization algorithm determines the best parameters of a mixture model for a given number of components. Since, in the general case, the number of components is not  known  beforehand, an extra loop of optimization is required to compare several optimal models with different number of components. To perform this task, we adopted the Akaike information criterion (AIC) as a cost function to asses the relative fitting quality between different proposed mixtures. In the case of the GES and APOGEE samples, the AIC gave preference for a two-component solution with  centroid and width values as quoted in each panel. The narrow metal-rich component is quite similar in both datasets, while the metal-poor one is broader and relatively more-metal poor in the  GES data.

The ARGOS MDF was found in \citet{ness2013} to be composed by up to five metallicity components, with the three metal-richest (designated as A, B and C: $\rm [Fe/H]>-1.0~dex$) accounting for the majority of stars. In that work, three general MDFs were assembled from samples in fields located in three latitude stripes ($\pm15^\circ$ in longitude) at  $\rm b=-5^{o},~-7.5^{o}~ \textmd{and} -10^{o}$. Their  GMM analysis yielded a three-component solution in each case. In particular, in the lower panel of Fig.~\ref{fig:MDF}, we display in light blue the three Gaussian components resulting from their analysis of the $\rm b=-5^{o}$ stripe. For verification, we performed a GMM analysis on the m0m5 field sample, which is depicted by the histogram. The best model is a mixture of three Gaussians with means of $\mu=-0.69,\ -0.27,\ 0.14$ and dispersions of $\rm \sigma=0.14\,dex,\ 0.16\,dex,\ 0.11\,dex$, in good agreement with the \citet{ness2013}  results for the entire  $b=-5^{o}$ strip, as quoted in the lower panel of Fig.~\ref{fig:MDF}.

The trimodal metallicity distribution of ARGOS data is visible in the individual field distributions as well as in the merged strip samples. This feature might be an imprint of the parametrization performed on the ARGOS data, but also, it could arise as an effect of assembling samples over a large longitudinal area, where small systematic variations in the intrinsic shape of the bulge MDF are possible.  In particular, the trimodal nature of the ARGOS parametrization could come from the fact that ARGOS uses effective temperature estimates from $\rm (J-K)_{0}$ colours (\citealt{freeman2013}), which can result in systematic differences (see Fig.~\ref{DeltaTeff} ) compared to spectroscopically-derived temperatures. \citet{recio-blanco16} estimated the end-of-mission expected parametrization performances of the Gaia DPAC pipeline (GSP-Spec) used for the derivation of the atmospheric parameters and chemical abundances from the RVS stellar spectra (R=11,200). The spectral resolution as well as the spectral region of the Gaia-RVS is very similar to that of ARGOS. \citet{recio-blanco16} show,  based on model spectra (see their Fig.~21), that an error of 100\,K can result easily in at least  0.1\,dex errors in [Fe/H] with a significantly large dispersion that gets larger for cooler stars. An effect such as this  could be the reason for the  additional third component in the observed MDF of ARGOS. 
On the other hand, the bimodality of the bulge MDF has been characterized by a number of studies examining specific locations in the bulge region \citep{uttenthaler2012,gonzalez2015}, as well as in the results of the fourth internal data release of the GES (Rojas-Arriagada et al. 2017), which covers a larger area.
To understand the difference between the ARGOS data set and those of APOGEE and GES, a common set of observed stars covering the full metallicity range -which is currently not available- is necessary.
In the end, this discrepancy highlights that understanding the number of metallicity-distinguished components constituting the bulge is an important and unresolved issue warranting further investigation.

\begin{figure}[!htbp]
{\center \includegraphics[width=0.48\textwidth,angle=0]{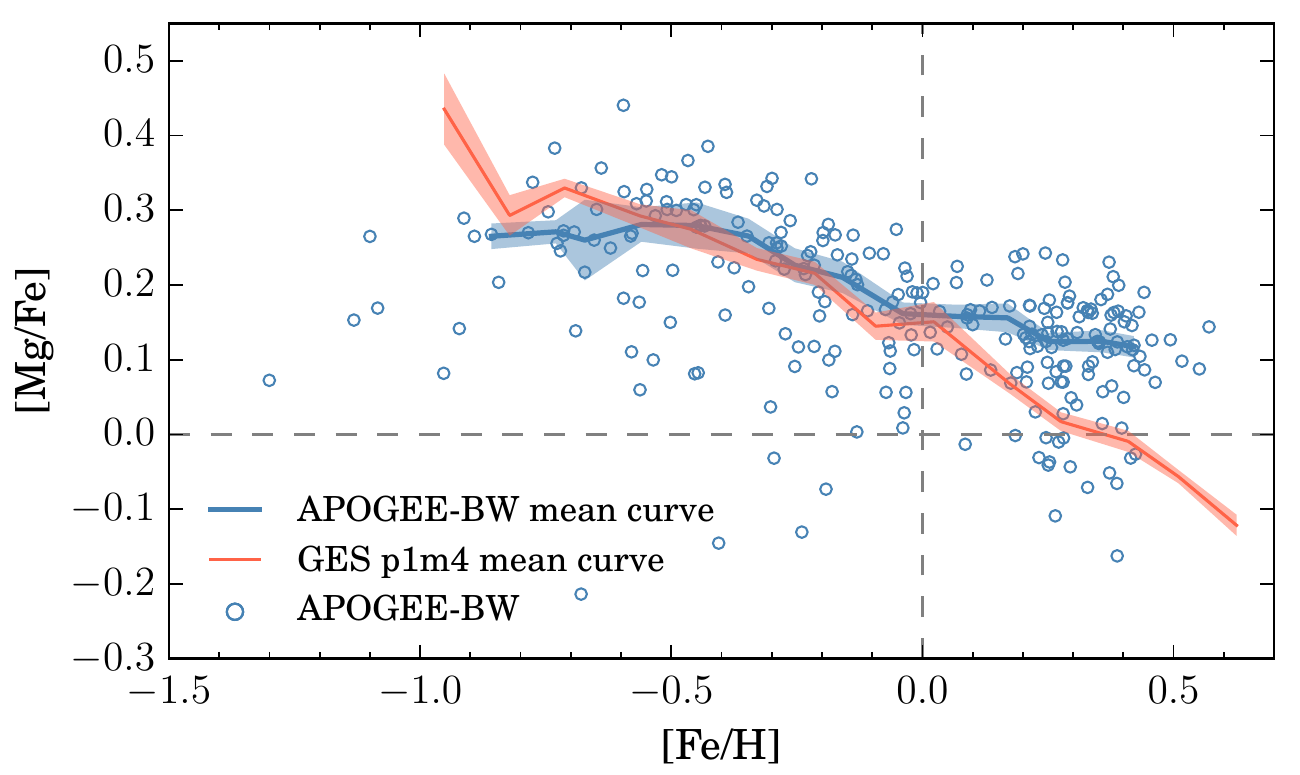}}
\caption{ Baade's window stars in the [Mg/Fe] vs. [Fe/H] plane. The APOGEE sample and its median trend (blue open circles and blue line, respectively) are compared with the mean trend of BW stars from the fourth internal data release of the GES (red line). A shaded red area around the mean trend depicts the standard error of the mean.} 
\label{fig:apogee_vs_ges_profile}
\end{figure}

\section{Trends in the abundance-metallicity plane}\label{sec:abundancetrends}
The distribution of stars from a given population in the abundance-metallicity plane encodes important information about its star-formation history and chemical evolution. In particular, detailed comparisons between the bulge and other Galactic components in the [$\alpha$/Fe] vs. [Fe/H] plane can provide a direct means of unraveling  the origin of the bulge in the context of other Galactic stellar populations.

Fig.~\ref{fig:apogee_vs_ges_profile}  displays our BW APOGEE sample in the [Mg/Fe] vs. [Fe/H] plane. To compare with the  GES, we overplot the median trend of its BW  stars,  as determined by dividing the sample into narrow metallicity bins (Rojas-Arriagada et al. 2016). To assess  the statistical significance of the resulting profile, a shaded area depicts the standard error of the mean. To enhance the comparison, we computed in the same manner the median trend of the bulge APOGEE sample studied here. The abundance scale of APOGEE agrees remarkably well with that of the  GES. The GES stars are on average slightly less $\alpha$-enhanced than APOGEE stars, by about 0.02\,dex over the common metallicity range. There is a clear discrepancy between the $\alpha$-element abundances between APOGEE and GES for the more metal-rich stars ($\rm [Fe/H] > -0.1$), in the sense that the GES obtains lower $\alpha$-element abundances. The difference is $\rm \sim 0.1\,dex$ at [Fe/H]=+0.2 and increases  to $\rm \sim 0.15\,dex$ at [Fe/H]=+0.4\,dex.
  APOGEE derives Mg abundances from four different spectral windows (centered at 1.533, 1.595, 1.672, 1.676 $\rm \mu$m), all of them in the infrared H band (see \citealt{garcia16}), while GES uses the lines at  8717.8 \AA,  8736.0 \AA, and 8806.7 \AA\,  in the optical spectral range.  A part of the differences between APOGEE and GES could arise from differences in atomic data, although one has to be aware that
abundance determination for metal-rich giants, as those used in these two studies, is challenging and systematic effects from other sources would not be surprising.

 This comparison emphasizes the necessity of a common sample of stars well-distributed in the $\rm T_{eff}$-$\rm \log(g)$-$\rm [Fe/H]$ space  to cross-calibrate abundance measures coming from different surveys.

\begin{figure}[!htbp]
{\center \includegraphics[width=0.49\textwidth,angle=0]{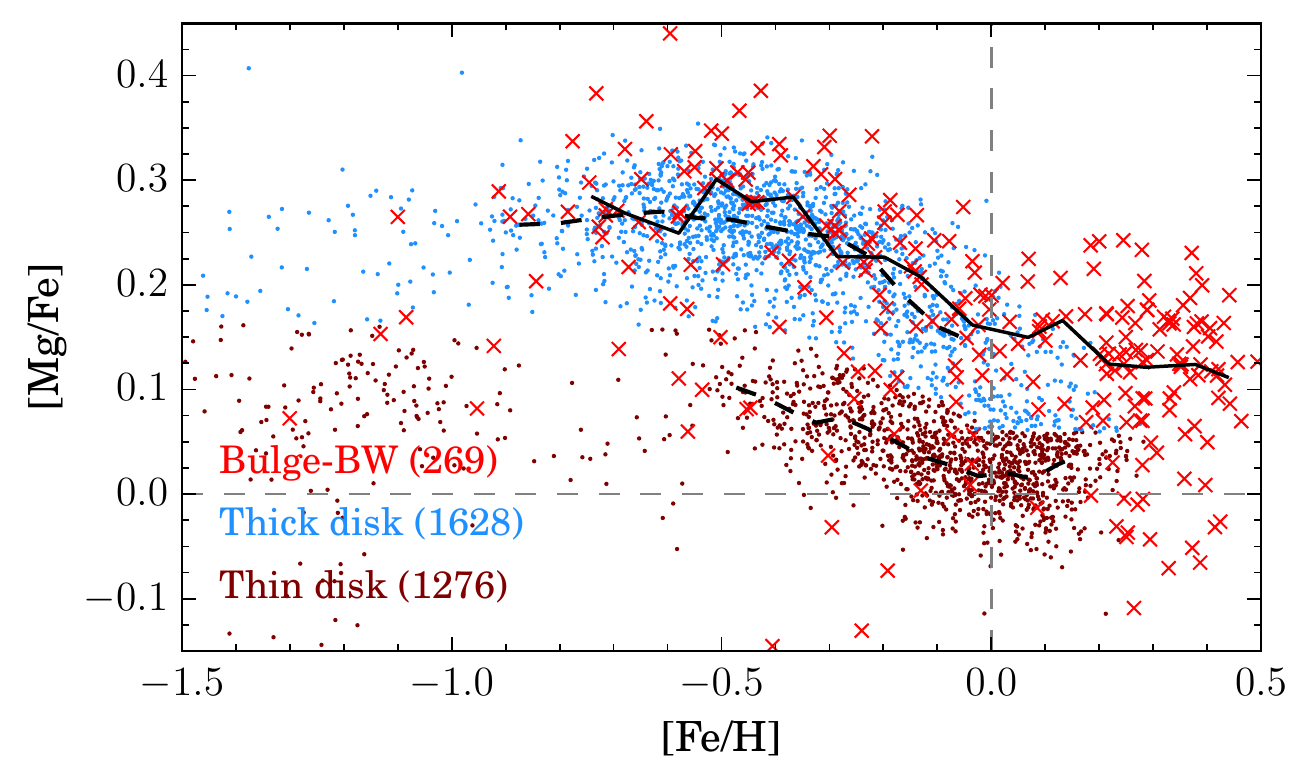}}
\caption{ [Mg/Fe] vs. [Fe/H] for thin disk (brown points), thick disk (blue points), and BW (red crosses) stars. Median trends are computed for the disks (dashed lines) and the bulge (solid line) in small metallicity bins, as a  visual aid. Bulge stars appear to be on average more $\alpha$-enhanced than the thick disc.}
\label{fig:disk_vs_bulge_apogee}
\end{figure}

In Fig.~\ref{fig:disk_vs_bulge_apogee} we attempt a direct homogeneous comparison between the distributions of disc and bulge stars in the abundance-metallicity plane, using results exclusively from APOGEE. To this end, we selected a sample of disc stars (namely, stars in pointings with $|l|\geq15^{o}$ and $|b|\geq15^{o}$), cleaning it based on several flags provided by the ASPCAP pipeline. In addition, we selected stars with $R_{GC}<7.7$~kpc to have a sample representing the chemical distributions inside the Solar circle, without decreasing significantly the sample size. To minimize the  systematics arising from the comparison of stars with different fundamental parameters, we selected disc stars in the same range of $\rm T_{eff}$ and $\rm \log\,g$ spanned by our BW sample. Application of all of  the previous cuts yield a final sample of 2904 stars. 

The selected disc stars present a distribution in the abundance-metallicity plane, with a clear gap separating a high-$\alpha$ and low-$\alpha$ sequences (see also \citealt{hayden15}). We divide the sample  by performing a clustering analysis in narrow metallicity bins. The resulting thin- and thick-disc samples (brown and blue points, respectively), together with the BW bulge sample (red crosses), are depicted in Fig.~\ref{fig:disk_vs_bulge_apogee}. To aid visualization of   their distributions, we compute a median trend for the discs (dashed lines) and the bulge (solid line). Interestingly, bulge stars appear to be on average more $\alpha$-enhanced than the thick disc, reaching  higher metallicity values. This may be the result of a difference in the relative formation timescales, with that of the bulge being faster and dominated by massive stars. The confirmation of this feature is of clear importance in our quest for disentangling the  different nature and origins of the stellar populations   that co-exist in  the central kiloparsecs of the Galaxy.

There are a small number of stars with low magnesium abundances. Their  relative proportion does not decrease significantly if we apply a more stringent cut in Galactocentric distances, which means that, if we account for the large errors in spectrophotometric distances, these stars appear to be located inside the bulge region. Recio-Blanco et al. (2016, submitted) found from GES Bulge data a small fraction of low-$\rm \alpha$-stars that have chemical patterns compatible with those of the thin disc, indicating a complex formation process of the galactic bulge

\section{The age distribution}\label{sec:age}

\begin{figure}[!htbp]
\centering
\includegraphics[width=0.40\textwidth,angle=0]{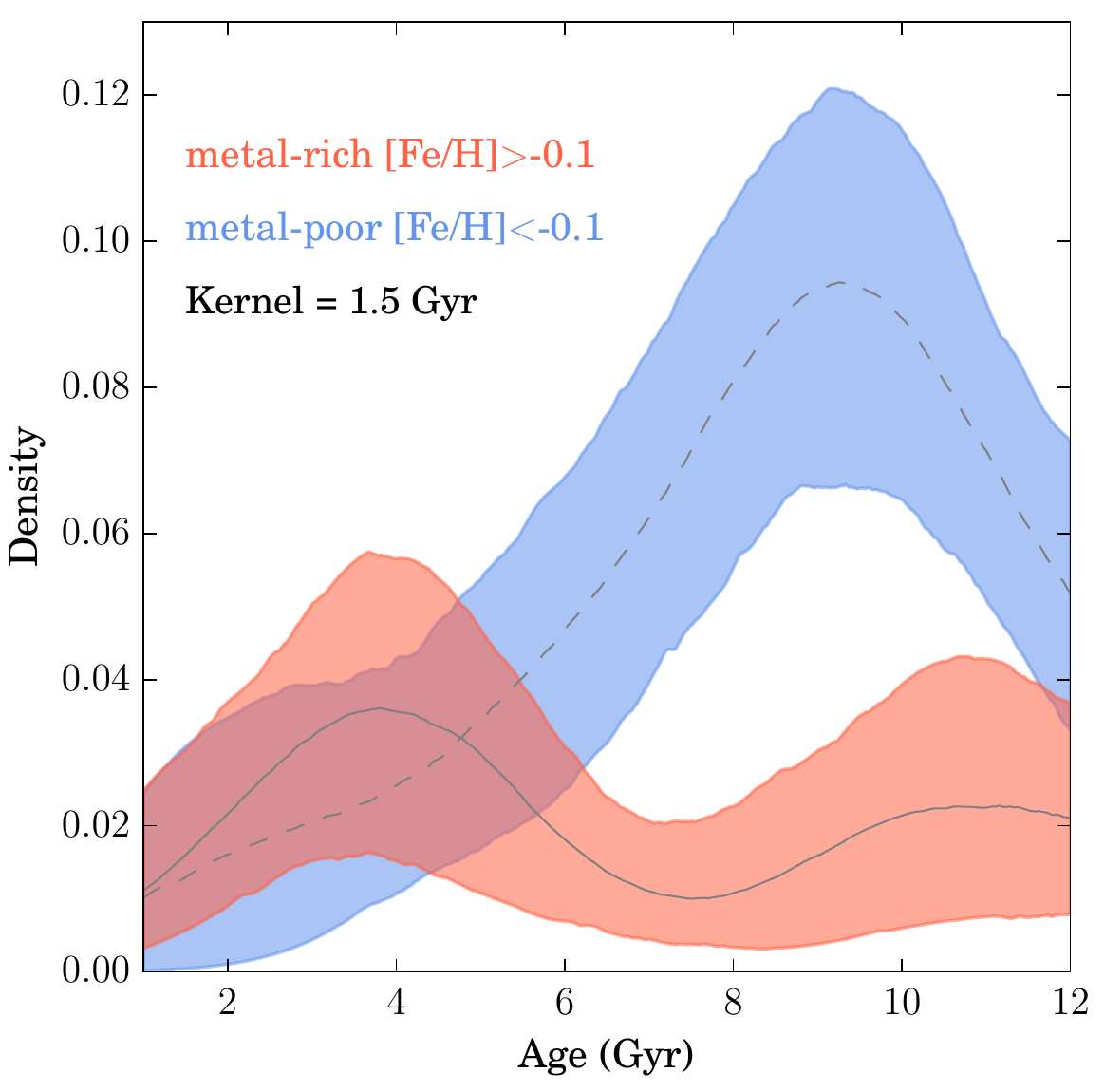}
\caption{Age distribution  of stars in BW using the formula of Martig et al. (2016).  Only sources with $\rm 4000 < T_{eff} < 5000\,K$, $\rm 1.8 < log\,g < 3.3$, $\rm [M/H] > -0.8$, $\rm -0.25 < [C/M] < 0.15$, $\rm -0.1 < [N/M] < 0.45$, $\rm -0.1 < [(C+N)/M] < 0.15$, $\rm-0.6 < [C/N] < 0.2$ were taken into account. From a bootstrap analysis on generalized histograms (kernel 1.5~Gyr), mean trends (solid and dashed gray lines) and errorbands to the $\pm2\sigma$ level (shaded color areas) are derived for metal-rich and metal-poor stars (cut at $\rm [Fe/H]=-0.1$~dex) as percentiles of the 600 bootstrap resamplings.}
\label{age}
\end{figure}

Obtaining accurate ages for stars in the Galactic bulge is a crucial ingredient in the comparison of observed data to chemo-dynamical evolutionary models. Recently, \citet{martig16} developed a new method to estimate masses and implied ages for giant stars, based on C and N abundances calibrated on asteroseismic data. They demonstrate that the $\rm  [C/N]$ ratio of giants decreases with increasing stellar mass, as expected from stellar-evolution models. We use the relation of \citet{martig16} from Appendix~A.3, and adopt the same cuts as those authors to ensure the reliability of the relation:  $\rm 4000 < T_{eff} < 5000\,K$, $\rm 1.8 < log\,g < 3.3$, $\rm [M/H] > -0.8$, $\rm -0.25 < [C/M] < 0.15$, $\rm -0.1 < [N/M] < 0.45$, $\rm -0.1 < [(C+N)/M] < 0.15$, $\rm-0.6 < [C/N] < 0.2$. This leaves only 74 stars; Fig.~\ref{age} shows the age distribution of stars with metallicity lower and higher than ${\rm [Fe/H]=-0.1~dex}$\footnote{Results below do not qualitatively change if the cut is done at $\pm0.1$~dex from this limit.} (to roughly separate stars in the two modes of the MDF, see Fig.~\ref{fig:MDF}). Given the small size of the sample, we use generalized histograms (kernel of 1.5~Gyr) to avoid the effect of binning of conventional histograms, and a bootstrap analysis to asses for the significance of the resulting distributions. To this end, we performed 600 bootstrap resamplings of the metal-rich and metal-poor samples, computing median trends (solid and dashed gray lines) and errorbands (shaded areas) at the $\pm2\sigma$ level from percentiles.

The peak of the distribution of metal-poor stars is about $\rm \sim 10\,Gyr$, with a decreasing tail toward younger ages. This compares  well with the mean bulge age as estimated from photometric data (\citealt{zoccali2003}, \citealt{clarkson08}, \citealt{valenti2013}). On the other hand, the generalized distribution of metal-rich stars shows a flatter distribution, with two overdensities of young and old stars. This seemingly bimodal age distribution for metal-rich stars is comparable with the results of \citet{bensby2013}, who found from their sample of dwarf and subgiant microlensed stars that while metal-poor bulge stars are uniformly old, metal-rich bulge stars span a broad range of ages (2-12~Gyr), with a peak at 4-5~Gyr. Recently, \citet{haywood16} concluded, from deep HST data in the  SWEEPS field , that a certain fraction of young stars are necessary to reproduce the observed colour-magnitude diagram (CMD). In their model, about 50\%  of the stars have ages greater than 8\,Gyr, suggesting that there might be a fraction of young stars in their CMD. If we extraploate their results to BW's, we would expect according to their model having 35\% stars younger than 8\,Gyr. We find a  very  similar fraction as seen in Fig. ~\ref{age}.  
However, their model reports that metal-rich stars with $\rm [Fe/H] > 0.0$ are all younger than 8\,Gyr, while our small sample suggests that metal-rich stars can be either young or old. We also note that the younger population exhibit, on average, less $\alpha$-element enhancement compared to the old population ($\rm alpha_{mean} = 0.126$ for ages $\rm < 6\,Gyr$ and $\rm alpha =0.215$ for ages $\rm > 6\,Gyr$ ). Overall, our age distributions derived from chemistry seem robust despite of the sample size, constituting an independent verification of results suggested from isochrone fitting to fundamental parameters \citep{bensby2013}, and photometric data \citep{haywood16}. However, we want to stress that the ages derived from [C/N] abundances have to be considered with caution. As discussed by \citet{martig16},  the absolute scaling of the derived  ages might be slightly off. This leads e.g. to underestimated ages of old stars as shown in their Fig.~11.  In addition, due to our small sample of stars,  clearly more data are necessary to better constrain the bulge age distribution and its metallicity dependence.

\section{Individual chemical abundances}
Compared to the 15 individual elemental abundances determined in DR12 (\citealt{holtzman2015}), the DR13 results include seven new elements: P, Cr, Co, Cu, Ge, Rb, and Nd; DR13 in total includes elemental abundances for the 22 elements C, N, O, Na, Mg, Al, Si, P, S, K, Ca, Ti, V, Cr, Mn, Co, Ni, Cu, Ge, Rb, Y, and Nd. However, we only discuss the abundances for  11 of these elements in this section. The  reasons for exclusion of the other 11 elements are :

\begin{itemize}
\item We do not include C and N abundances as a giant star which ascends the giant branch, deepens the convective envelope and the star experiences the first dredge-up. This means that CNO-processed material containing a lot of N, but is depleted in C is brought to the surface. This is nicely shown in Fig.~1 of  Martig et al. (2016). It also turns out that the depth of the convective envelope as well as the amount of CNO-cycling in the core depends on the mass of the star, a fact that we use to get ages for the stars in our Sect. 7.  This means that giants cannot be used to trace the galactic chemical evolution of C and N: the abundances simply do not reflect the abundances of their birth. This is shown in the lower panel of Fig. 2 in Martig et al. (2016).
\item The spectral lines used to determine the P abundances are generally very weak in the type of giants observed in BW. Holtzman et al. (2017, in prep.)  caution that these lines are weak and uncertain, while \citet{hawkins16} derive only upper limits in their independent analysis of APOGEE spectra. We therefore exclude this element in the discussion below.
\item The abundance-trend of S for the BW stars is very scattered as compared to our sample of local disk stars. Possibly this is because the S abundance is in principle derived from a single, blended line, and higher S/N than the already high S/N of the BW spectra are needed to trace this element with certainty. We therefore exclude this element in the discussion below. 
\item The DR13 abundance trends of Ti in the local discs does not  resemble the expected $\alpha$-element  trends found in many other works. The reason for this behavior is described in  \citet{hawkins16}, as  possible 3D/NLTE-effects in the Ti I lines used in DR13. We therefore exclude this element in the discussion below.
\item The abundance trend of V in the BW stars is very scattered. The V abundances are mainly determined from two lines, of which one is quite weak, but the other one is of suitable strength. It is possible that the V abundance trend in the bulge would be less scattered if the weak line were to be excluded in the analysis. We also note that \citet{hawkins16}  derive a different V trend for their independent analysis of a sub-sample of APOGEE-spectra. For these reasons, we exclude this element in the discussion below.
\item The Cu, Ge, Rb, Y, and Nd abundances are all determined from few, often single, weak and blended lines.  We therefore exclude these elements in the discussion below. 
\end{itemize}

To conclude, we discuss the abundances of the following 11 elements in this section: O, Na, Mg, Al, Si, K, Ca, Cr, Mn, Co, and Ni.


\begin{figure*}[!htbp]
\centering
\includegraphics[width=0.35\textwidth,angle=270]{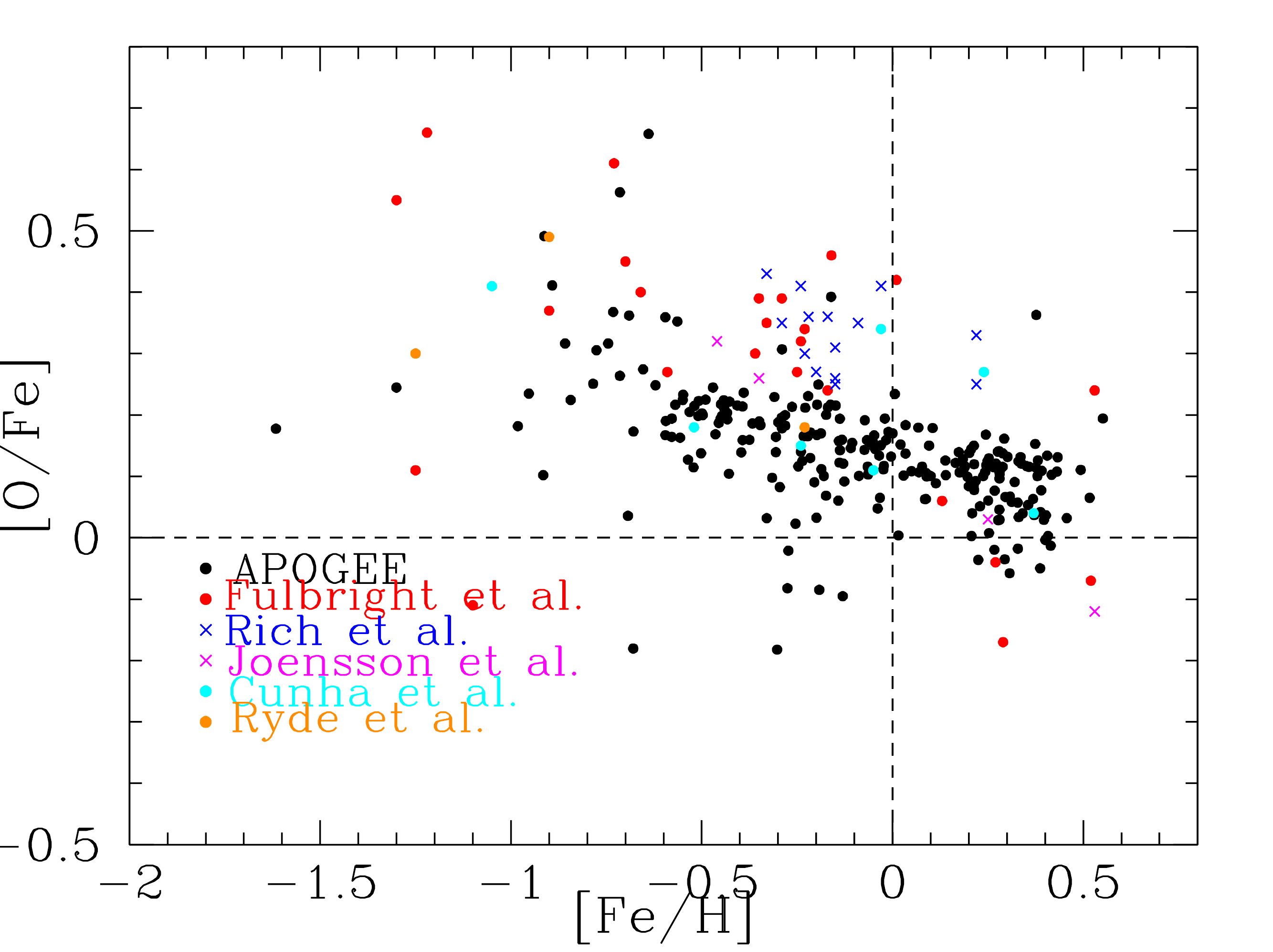}  \includegraphics[width=0.35\textwidth,angle=270]{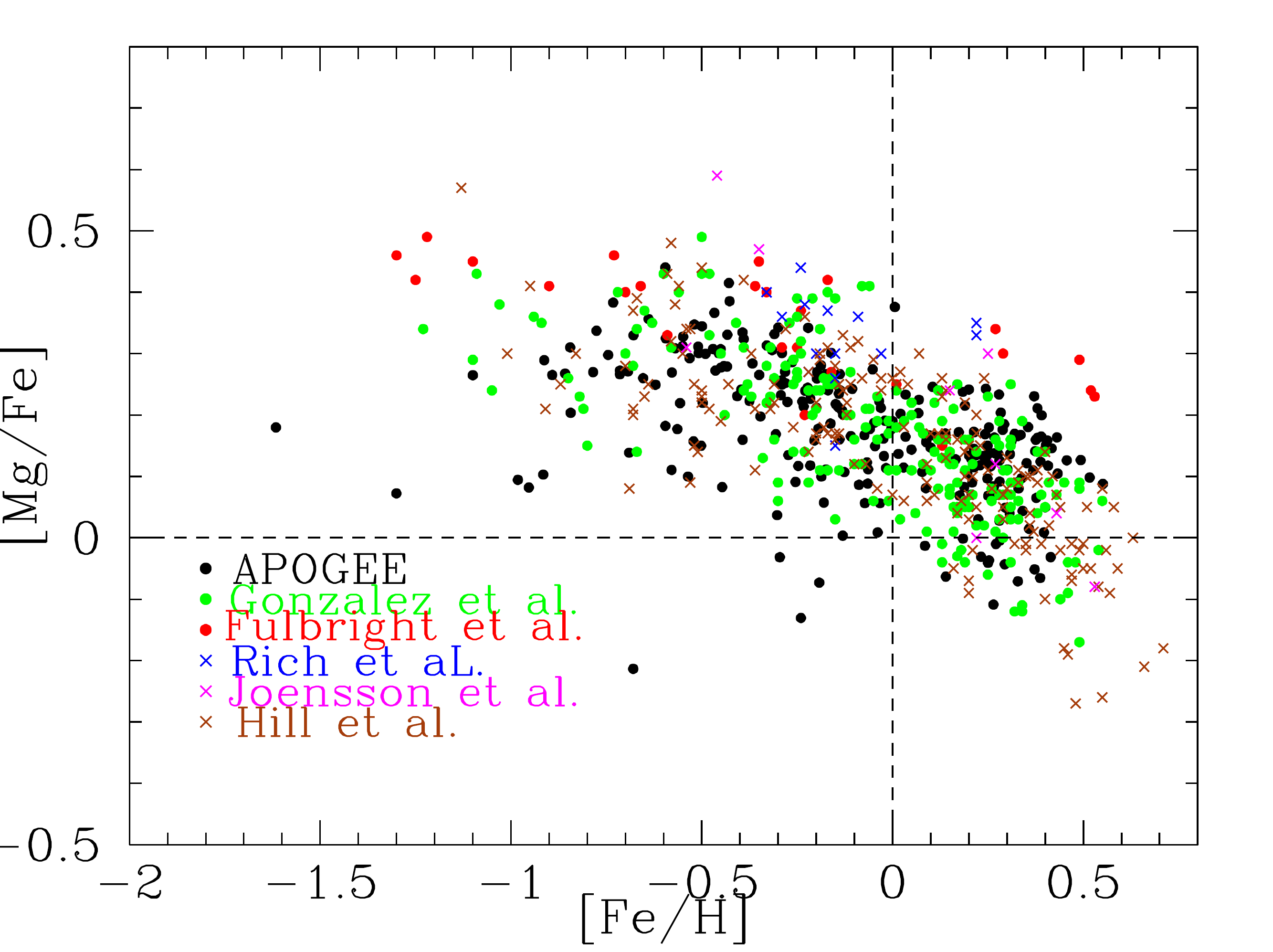}
\includegraphics[width=0.35\textwidth,angle=270]{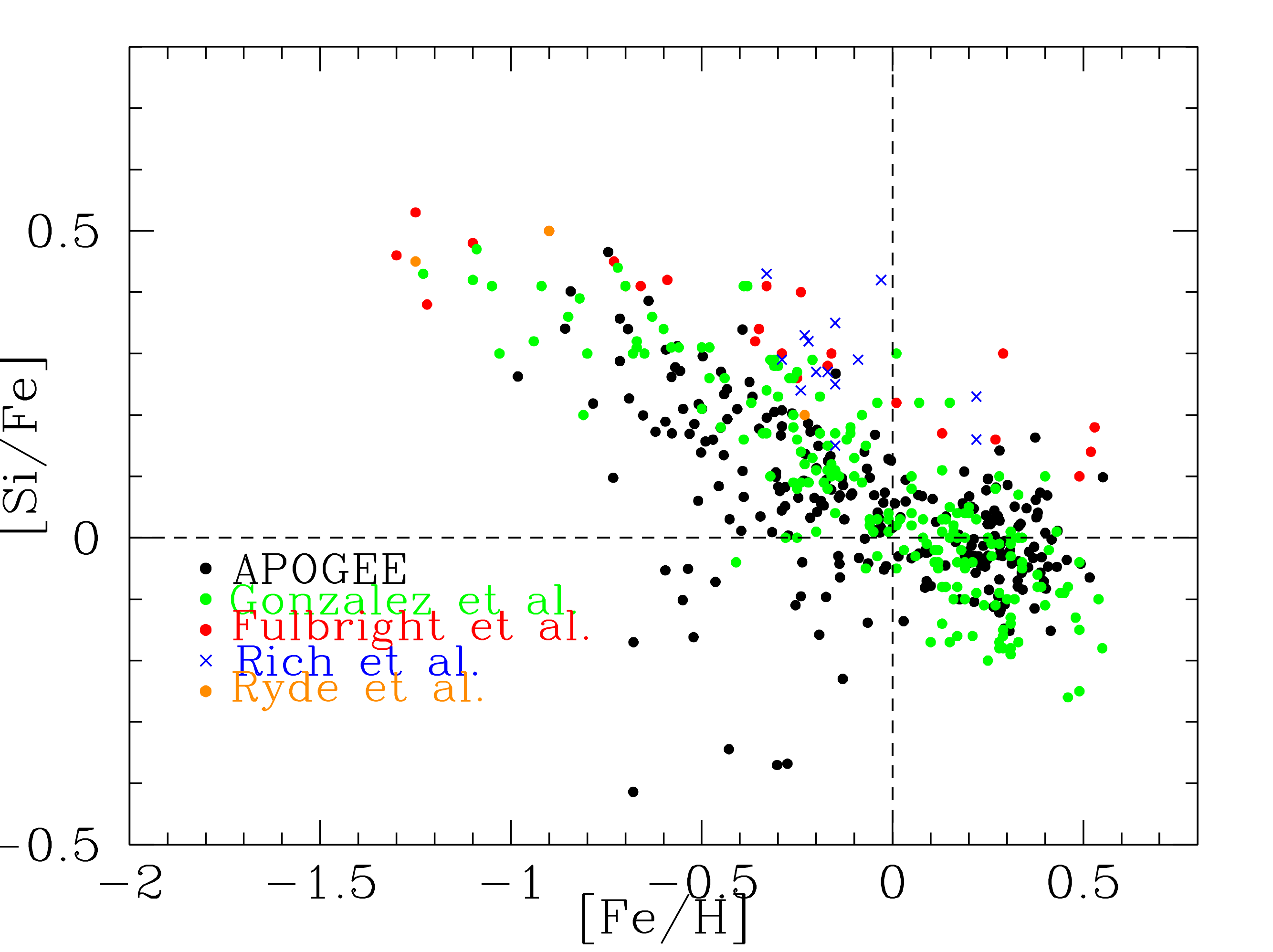} \includegraphics[width=0.35\textwidth,angle=270]{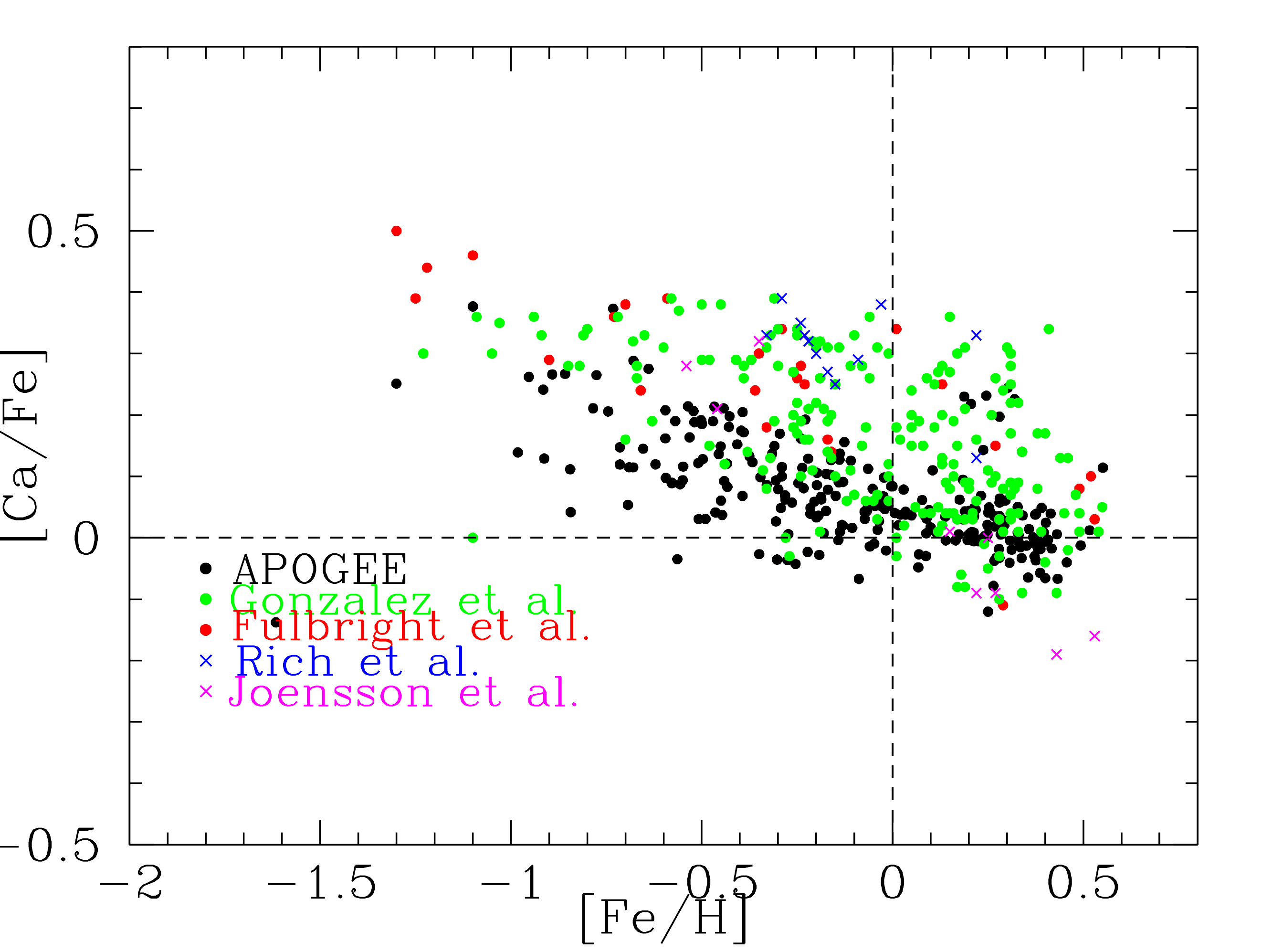}
\caption{[Fe/H] vs. [X/Fe] of APOGEE stars in BW for  $\alpha$-elements compared with literature values.}
\label{abundancesalpha}
\end{figure*}

\begin{figure*}[!htbp]
\centering
\includegraphics[width=0.35\textwidth,angle=270]{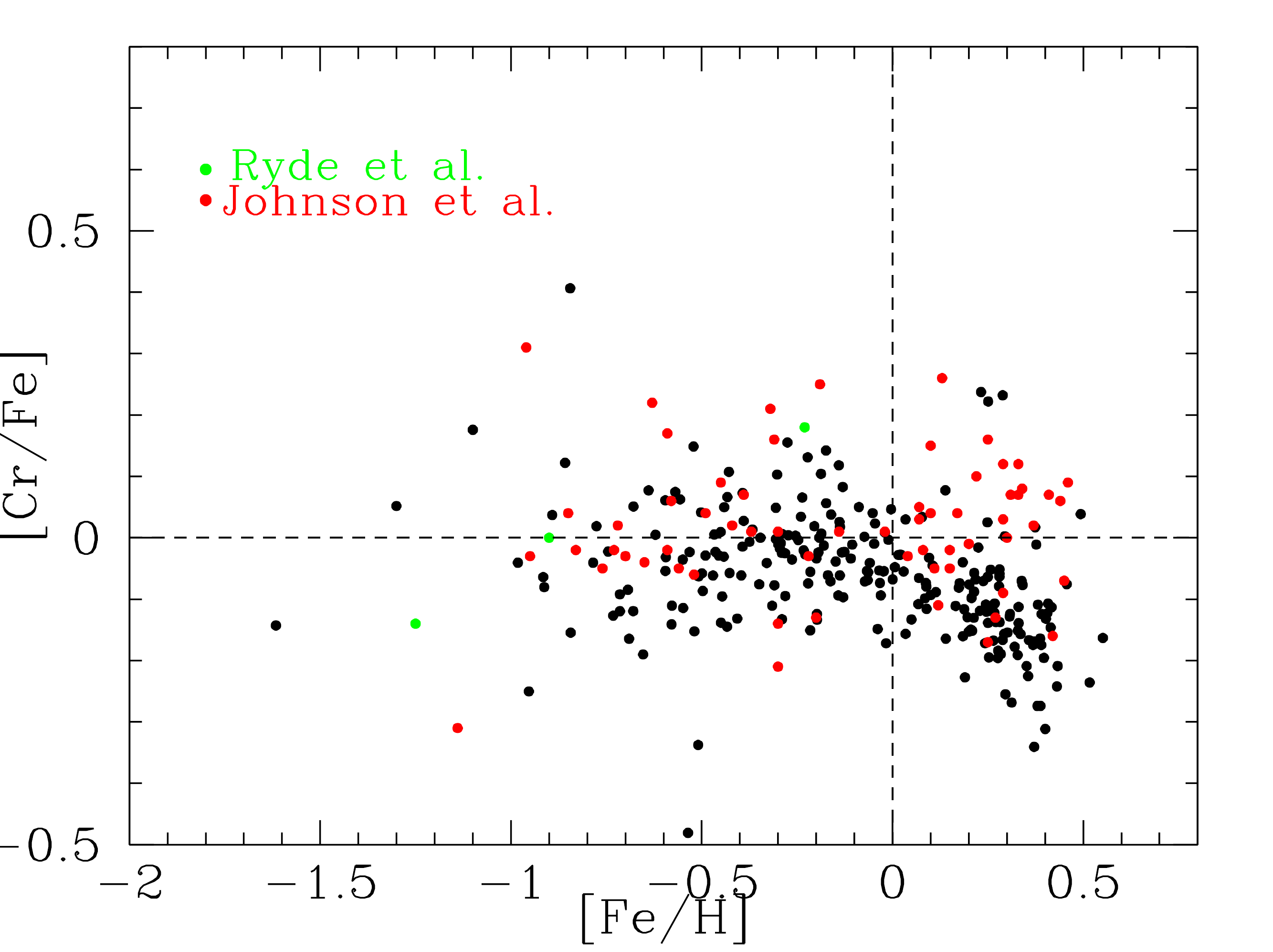} \includegraphics[width=0.35\textwidth,angle=270]{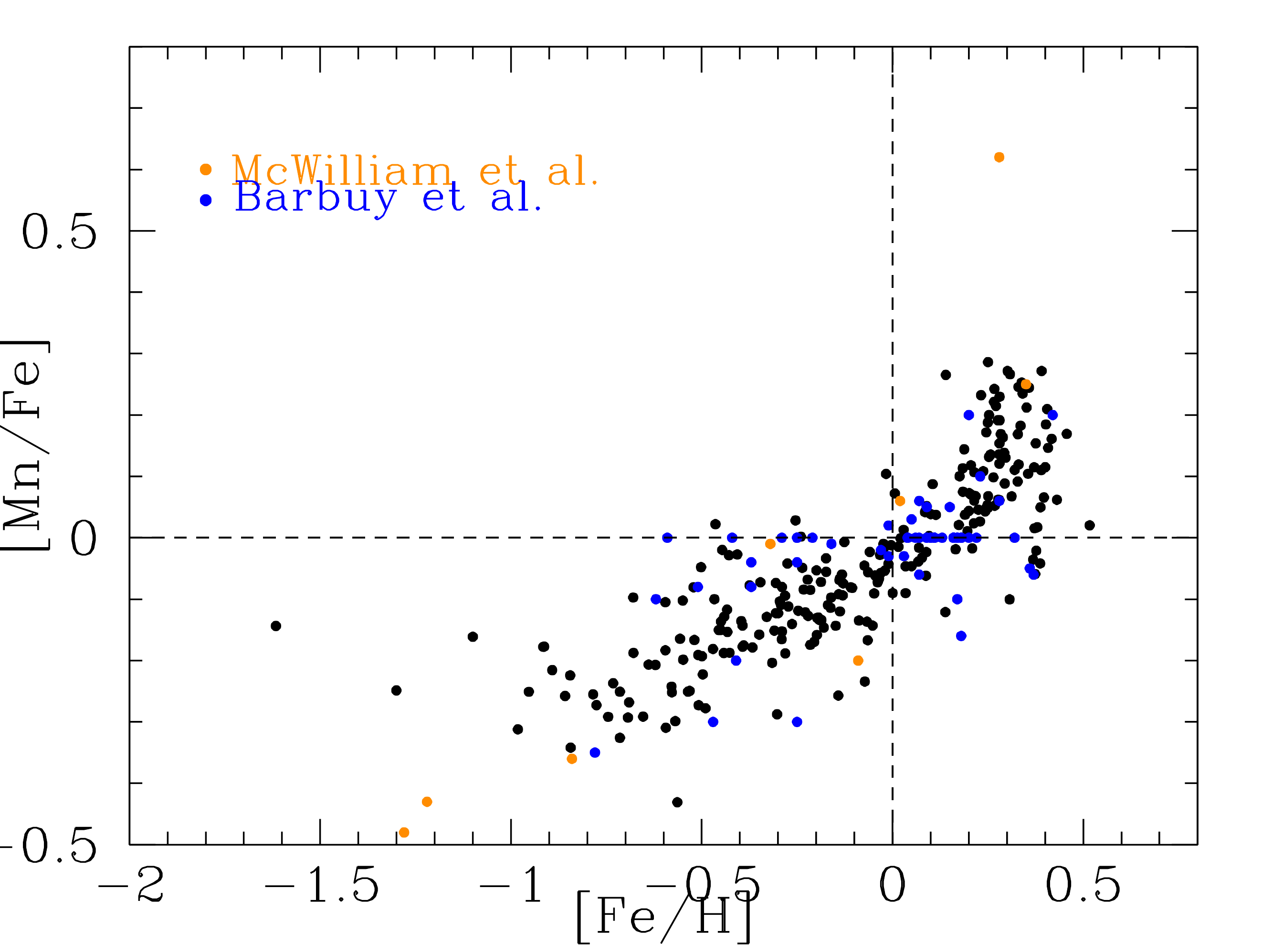}
\includegraphics[width=0.35\textwidth,angle=270]{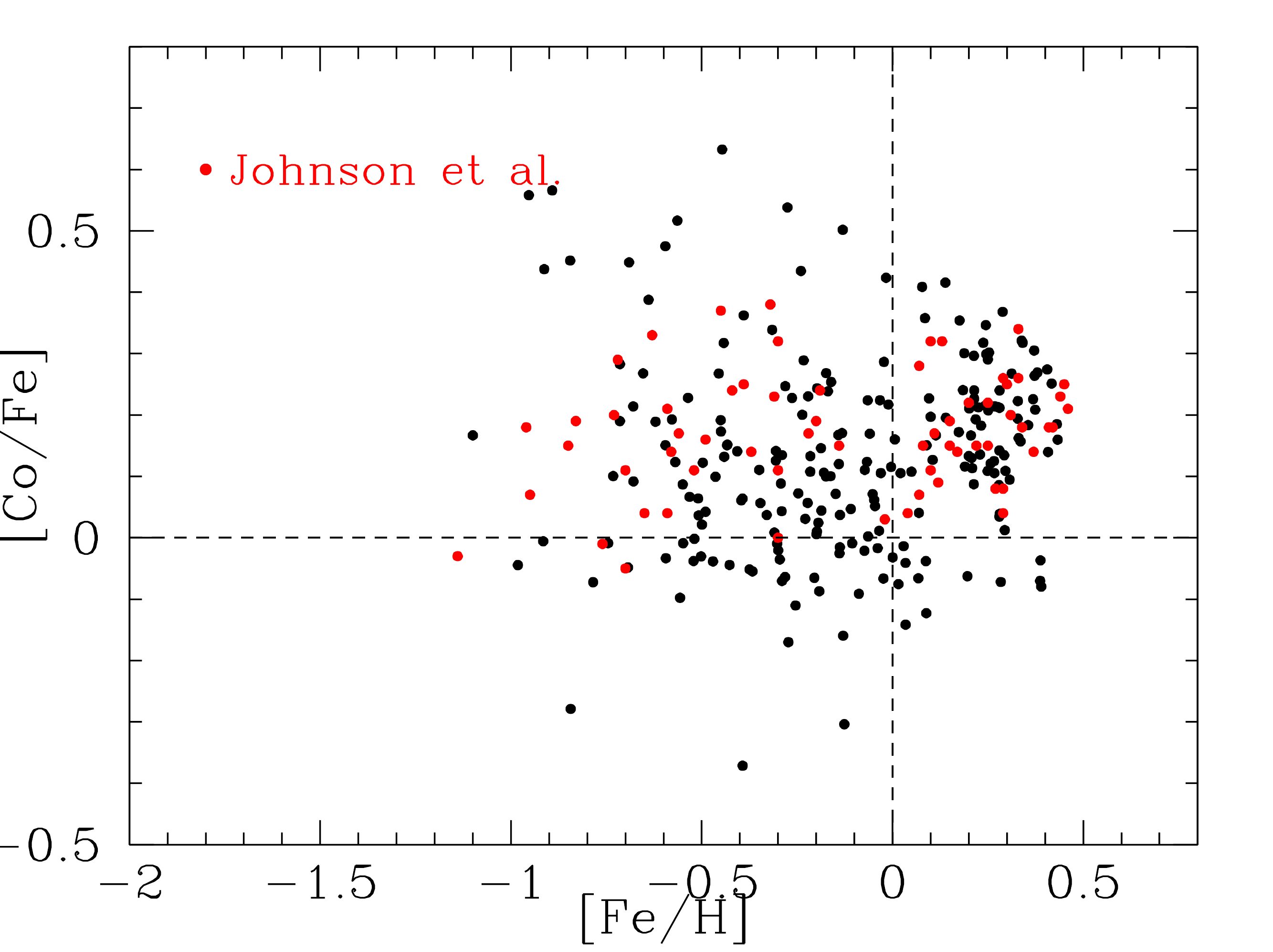} \includegraphics[width=0.35\textwidth,angle=270]{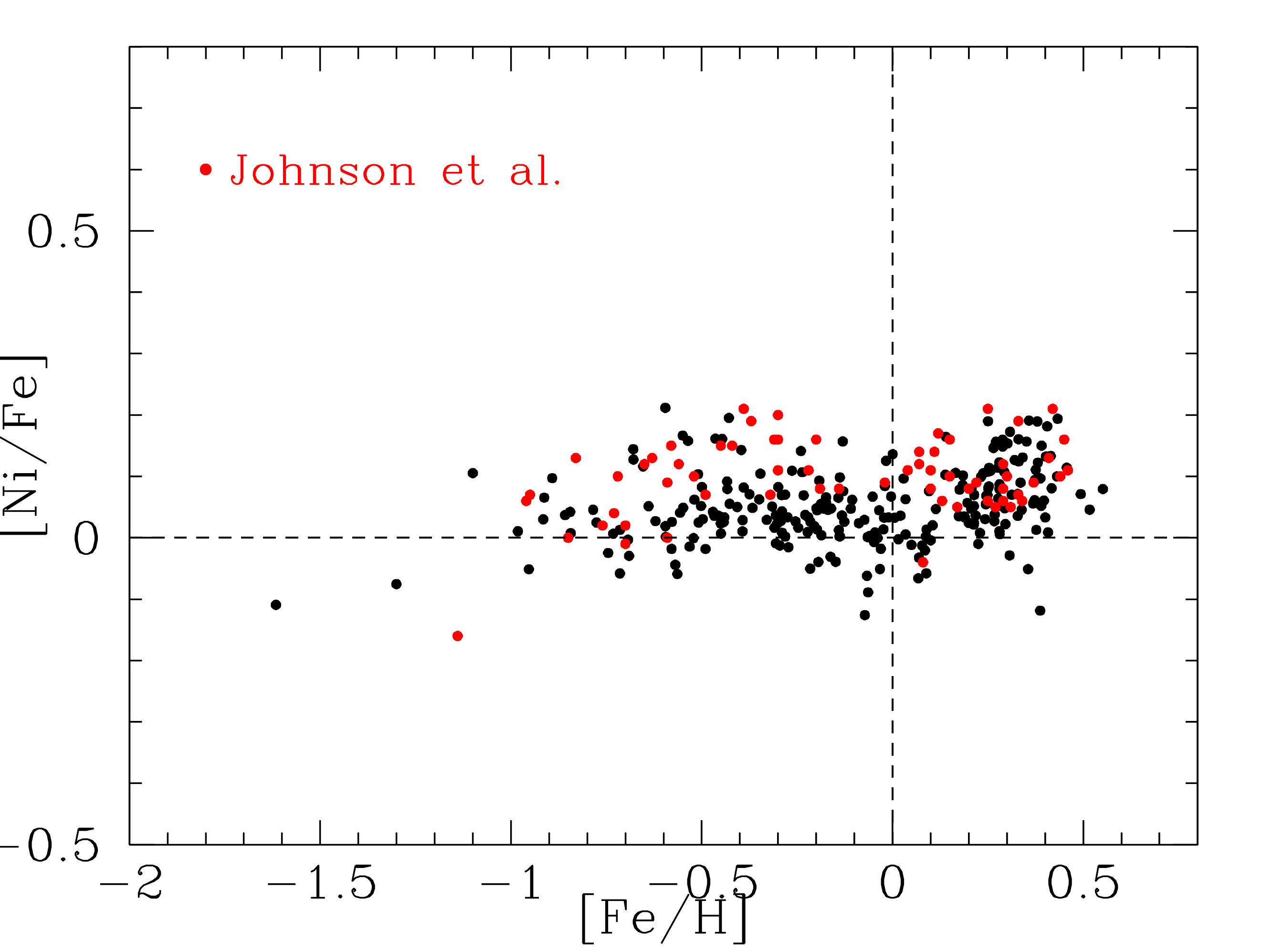}
\caption{[Fe/H] vs. [X/Fe] of APOGEE stars in BW for iron-peak elements compared with literature values.}
\label{abundancesironpeak}
\end{figure*}

\begin{figure*}[!htbp]
\centering
\includegraphics[width=0.35\textwidth,angle=270]{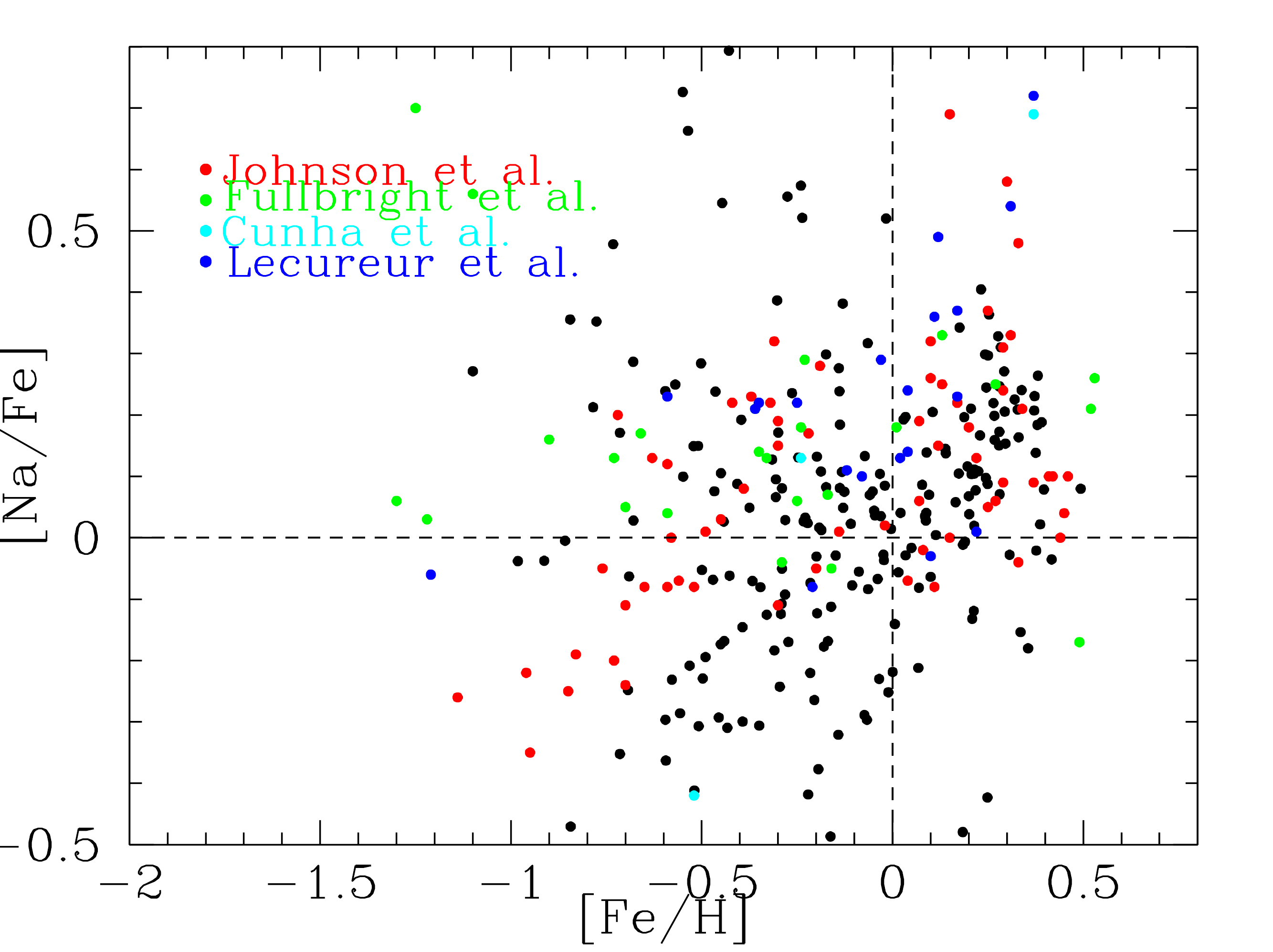} \includegraphics[width=0.35\textwidth,angle=270]{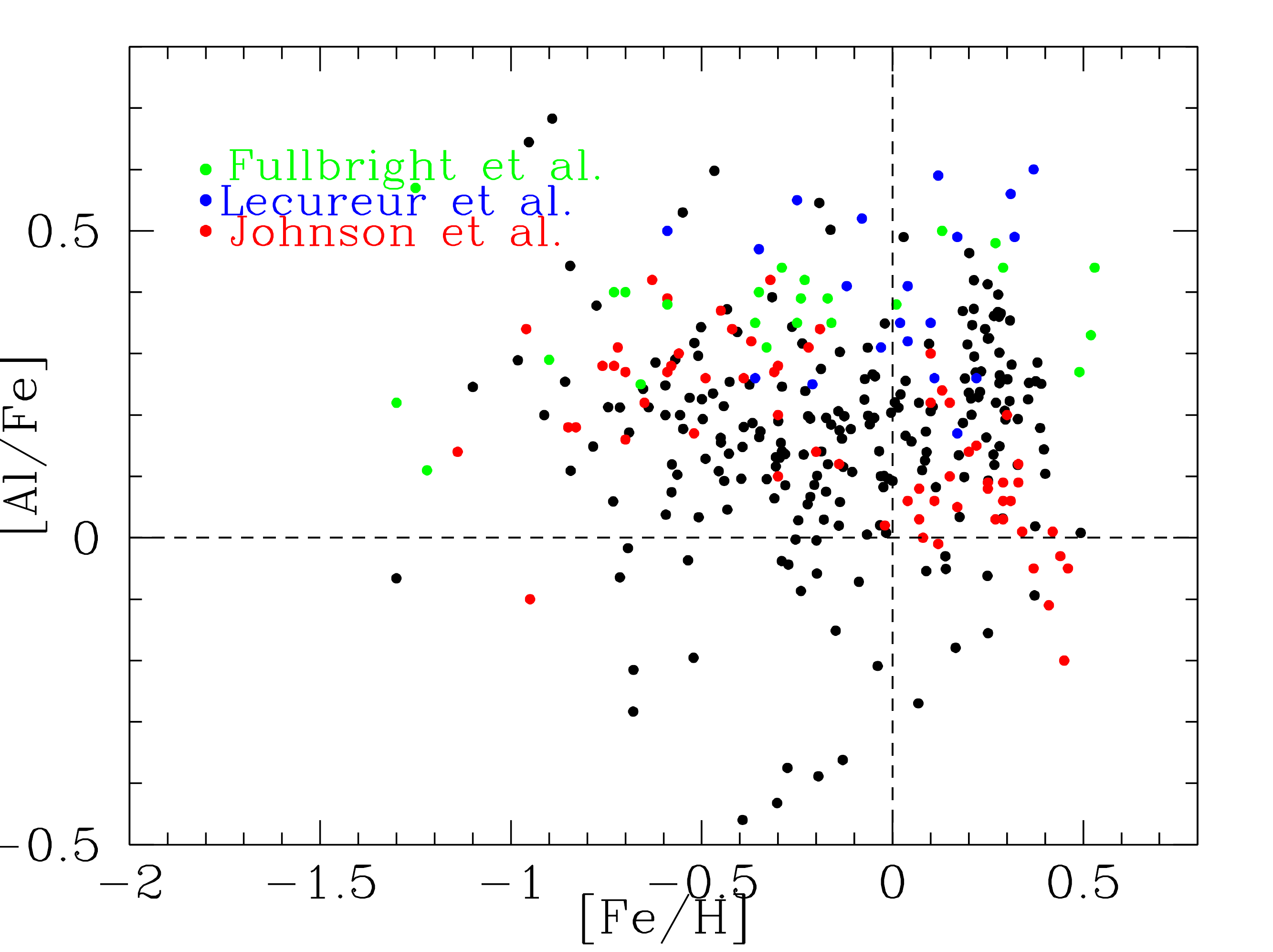}
\includegraphics[width=0.35\textwidth,angle=270]{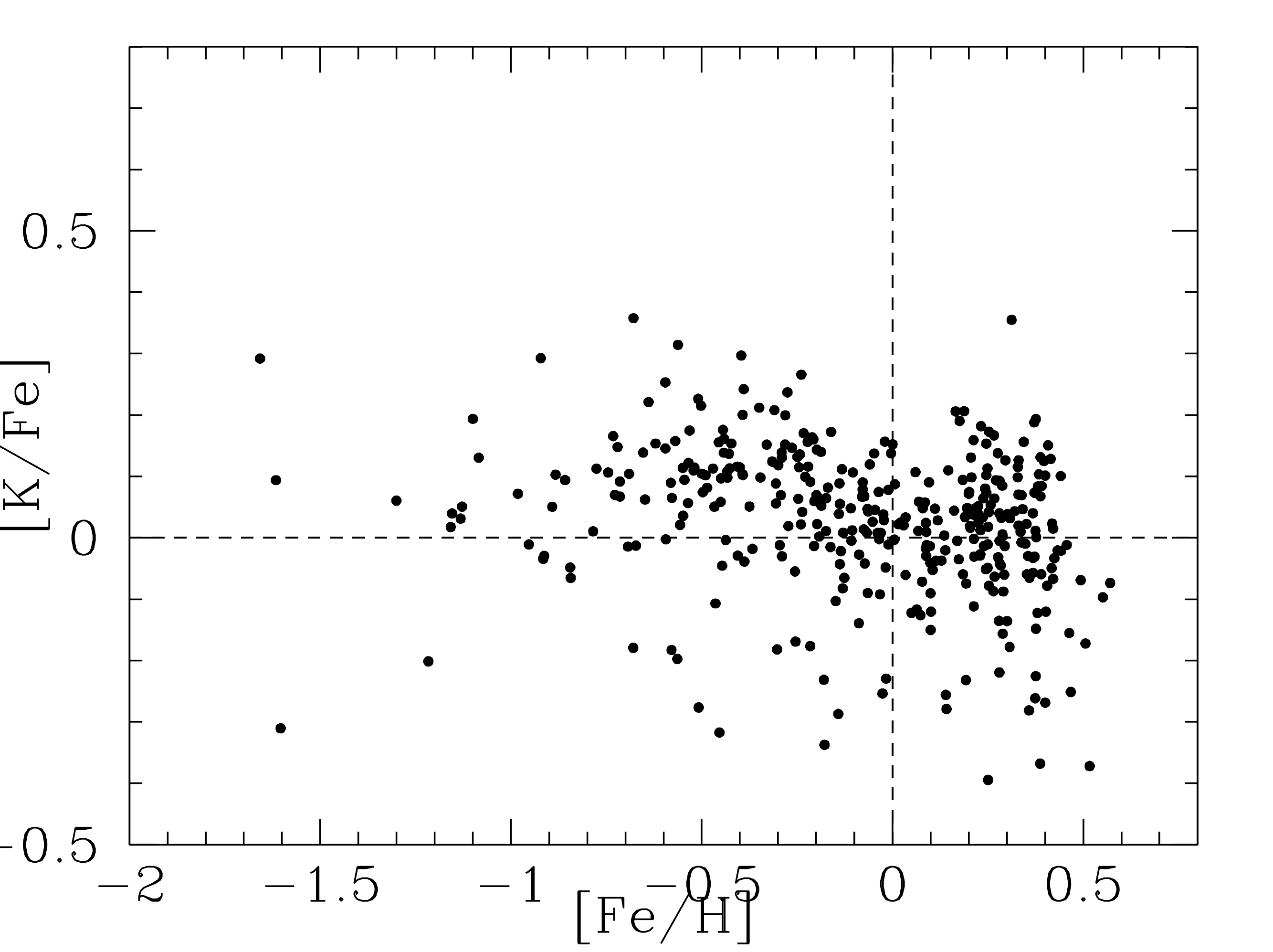}  
\caption{[Fe/H] vs. [X/Fe] of APOGEE stars in BW for odd-Z elements compared with literature values.}
\label{abundancesodd}
\end{figure*}

\subsection{The $\alpha$-elements: O, Mg, Si, and Ca}
Figure~\ref{abundancesalpha} shows the trends of the $\alpha$-elements (O, Mg, Si, Ca) compared to available literature values for M giants. These abundances are of particular interest, because accurate $\rm [\alpha/Fe]$ ratios place strong constraints on the star-formation history (e.~g. \citealt{matteucci1990}) in a stellar population. In addition to the previously mentioned references, we add that of \citet{jonsson2016}, who determined elemental abundances of O, Mg, and Ca of bulge K giants using the high-resolution ($\rm R \sim  47 000$) UVES/FLAMES spectrograph at the  VLT. 

Oxygen abundances determined in Rich05, Cunha06, \citet{fulbright07}, Ryde10, and \citet{jonsson2016}  are presented in the top left panel of Figure \ref{abundancesalpha}. The APOGEE O abundances are lower than those of Rich05 and \citet{fulbright07}, while they are comparable to those of Cunha06, Ryde10, and \citet{jonsson2016}. 

Magnesium abundances are determined in Rich05, \citet{fulbright07}, \citet{gonzalez11}, \citet{hill2011}, and \citet{jonsson2016};  these abundances are displayed in the top right panel of Figure \ref{abundancesalpha}. The APOGEE Mg abundances exhibit generally   good agreement with the trends found by \citet{gonzalez11} and \citet{hill2011}. The Rich05, \citet{fulbright07}, and possibly \citet{jonsson2016} stars are slightly enhanced in Mg compared to APOGEE, in particular in the metal-rich regime ($\rm [Fe/H] > 0.2$).

Silicon abundances are determined in Rich05, \citet{fulbright07}, Ryde10, and \citet{gonzalez11}, and all abundances are shown in the bottom left panel of Figure \ref{abundancesalpha}. The APOGEE Si abundances are in general agreement with Ryde10 and \citet{gonzalez11}, while Rich05 and \citet{fulbright07} derive systematically higher Si abundances. 

Calcium abundances are determined in Rich05, \citet{fulbright07}, \citet{gonzalez11}, and \citet{jonsson2016}.  Those abundances are plotted in the bottom right panel of Figure \ref{abundancesalpha}. The APOGEE Ca abundances are systematically about 0.15\,dex lower as compared to Rich05, \citet{fulbright07}, and \citet{gonzalez11}. \citet{jonsson2016} report similar low Ca abundances as APOGEE. The dispersion in APOGEE is much smaller than, e.g., \citet{gonzalez11},  for a given metallicity, resulting in a narrow Ca sequence for the metallicity range $\rm -1 < [Fe/H] < 0.5$.
 

\subsection{The iron-peak elements: Cr, Mn, Co, and Ni}
Unlike the lighter elements, the abundance patterns of Fe-peak elements in the Galactic bulge are not well-explored; and only few studies exist for comparison. \citet{johnson14} investigated chemical abundances of $\alpha$-elements and heavy Fe-peak elements such as Cr, Co, and Ni, for RGB stars roughly 1\,mag above the clump in the Galactic bulge, however  not in BW. Due to the  lack of comparison samples  for the iron-peak elements,  we have chosen to include this reference. \citet{mcwilliam2003} studied manganese in BW, and found that the bulge $\rm [Mn/Fe]$ trend is approximately the same as found in the solar neighborhood disc, and as well for halo stars,  and even follows the local $\rm [Mn/Fe]$ trend for metal-rich stars. \citet{barbuy2013} obtained Mn measurements of 56 red giants using the high-resolution FLAMES/UVES spectra for four Galactic bulge fields, and concluded that  the behaviour of $\rm [Mn/Fe] vs. [Fe/H]$ shows that the iron-peak element Mn has not been produced in the same conditions as other iron-peak elements, like Fe and Ni.

The production sites for these elements are uncertain, and the stellar yields of these elements remain under debate (\citealt{battistini15}). Mn and Co are believed to be produced mainly by explosive silicon burning in SNII (\citealt{woosley1995}),  while to a smaller extent in SNIa (\citealt{bravo12}).

Manganese  exhibits an increasing trend with increasing metallicity (top right panel of Figure \ref{abundancesironpeak}), both for our determinations and for the literature values. There is a well-defined [Mn/Fe] trend in the metallicity range $\rm -0.7 < [Fe/H] < +0.2$; while for the most metal-rich stars ($\rm [Fe/H] > 0.2$) the Mn abundances have a larger scatter at a given $\rm [Fe/H]$. \citet{battistini15} have shown that the Mn trends can change drastically if NLTE corrections are used,  resulting in  $\rm [Mn/Fe]$  becoming basically flat with metallicity. 
The increasing $\rm [Mn/Fe]$ trend is consistent with previous studies for the thick disc and halo stars in the range $\rm -1.5 < [Fe/H] < 0.0$ (\citealt{prochaska2000}, \citealt{nissen2000}). For $\rm [Fe/H]  > -1$, the increasing $\rm [Mn/Fe]$ with increasing $\rm [Fe/H]$ is interpreted as an onset of contribution from Type Ia SNe (\citealt{kobayashi2006}).

The behaviour of  $\rm [Co/Fe]$ vs. $\rm [Fe/H]$ is shown in the bottom left panel of Figure \ref{abundancesironpeak}. $\rm [Co/Fe]$ exhibits low-level variations as a function of $\rm [Fe/H]$ but is generally enhanced with $\rm [Co/Fe] = +0.15$, which is similar to that observed by \citet{johnson14}.

\citet{johnson14} reported that $\rm [Cr/Fe] = 0.0$ for the full metallicity range,  and that the abundance patterns of $\rm [Cr/Fe]$ are very similar to the thin-disc and thick-disc stars. The top left panel of Figure \ref{abundancesironpeak} reveals  similar $\rm [Cr/Fe]$ trends for $\rm [Fe/H] \le 0.0$, but contrary to \citet{johnson14}, our $\rm [Cr/Fe]$ trend decreases for $\rm [Fe/H] > 0$.

$\rm [Ni/Fe]$ displays similar variations to $\rm [Co/Fe]$, but at a much smaller amplitude,  and is slightly enhanced with $\rm [Ni/Fe] = +0.05$,  which is in good agreement with the results obtained by \citet{johnson14}.


 

\subsection{The odd-Z elements: Na, Al, and K}
Sodium abundances of bulge stars have been derived in Cunha06. Na and Al abundances of bulge stars are determined in \citet{lecureur07}, \citet{fulbright07}, and \citet{johnson14}. There is to our knowledge no previous  determination of K in bulge stars. The odd-Z elements Na and Al are believed to be produced in a variety of processes \citep[][and references therein]{smiljanic16} while  K is thought to be mainly formed in Type II SNe (\citealt{samland98}).
\citet{smiljanic16} reports evidence for the surface abundance of Na varying due to stellar evolution of giants, possibly making our sample of giant stars unsuitable to trace the chemical evolution of this element in the bulge. Figure \ref{abundancesodd} suggests, that,  compared to other bulge works using giants (Cunha06, \citealt{fulbright07}, \citealt{lecureur07}, and \citealt{johnson14}), our results have larger scatter, but are likely following the same  trend of rising [Na/Fe] for higher [Fe/H], as in \citet{fulbright07}, \citet{lecureur07}, and \citet{johnson14}.

Figure \ref{abundancesodd} also demonstrates  that our results appear to corroborate the aluminium trend found in \citet{johnson14}, albeit with a larger scatter. \citet{lecureur07} and \citet{fulbright07}, however,  find, on average,  higher values of [Al/Fe], especially  at higher metallicities. \citet{McWilliam2016} showed that [Al/Fe] displays an  alpha-like trend (see their Fig. 5) which is expected as Al production occurs in post carbon-burning hydrostatic phases of massive stars. A comparison of [Al/Fe] in the bulge, the Milky Way disk and Sgr dwarf galaxy suggests that the Al yields depend also on the progenitor metallicity (\citealt{fulbright07}).

Our [K/Fe] vs. [Fe/H] trend shown in Figure \ref{abundancesodd} observationally resembles an $\alpha$-like trend, with decreasing [K/Fe] for increasing [Fe/H], similar to  what APOGEE finds for the discs. However, the bulge [K/Fe]-values for the most metal-rich stars are higher than for the  disc stars of corresponding metallicity.

Concluding, we have  shown that chemical abundances APOGEE  in BW  (C, N, O, Na, Mg, Al, Si, K, Ca, Cr, Mn, Co, and Ni)  follow in general a tight sequence in the [Fe/H] vs [X/Fe] plane and agree well with known high-resolution abundance studies.

\section{Summary}

We have investigated the MDF for a large sample of stars  in BW with APOGEE, and found a remarkable agreement with the MDF of GES;  both exhibit a  bimodal distribution.  The ARGOS  survey, in contrast, exhibits  three distinguishable peaks.  The reason for this difference could be the use of photometric temperatures using (J--K) colours which could have effects on  the MDF.  In the [Mg/Fe]  vs. [Fe/H]  plane, APOGEE and GES  exhibit very similar results , although for higher  metallicities ($\rm [Fe/H] > +0.1$) APOGEE exhibits  higher abundance levels with respect to the GES.  We used  the [C/N] ratio to derive the age distribution for a subset of  the stars in BW, following \citet{martig16},  and found a bimodal distribution with a peak of $\rm \sim 10 \,Gyr$ and a significant fraction of  young stars ($\rm \sim 3-4\,Gyr$). Our findings are comparable with those of \citet{bensby2013} and \citet{haywood16}.  However,  more data are necessary to constrain the age distribution in BW.

We have compared stellar parameters and individual abundances for $\alpha$- and iron-peak elements from  the APOGEE pipeline (DR13) with known literature values for stars in BW. The difference between photometric log\,{g} values  and spectroscopic determinations shows a strong linear relation with the spectroscopic log\,{g} of APOGEE. TRILEGAL simulations suggest that this effect  is due to the intrinsic depth of the bulge, thus  photometric surface gravities in the Galactic bulge should be treated with caution. Compared to the relatively small number  of measurements in the literature, APOGEE traces  heavy elements covering a large metallicity range ($\rm -1 < [Fe/H] < +0.5$) in the bulge.
In general, the comparison between individual abundances of O, Na, Mg, Al, Si, K, Ca, Cr, Mn, Co, and Ni  from APOGEE  with that of literature values shows an overall  good agreement.

\begin{acknowledgements}
MS acknowledges the Programme National de Cosmologie et Galaxies (PNCG) of CNRS/INSU, France, for financial support. Henrik J\"onsson acknowledges support from the Birgit and Hellmuth Hertz' Foundation, the Royal Physiographic Society of Lund, Sweden. T.C.B. acknowledges partial support from grant PHY 14-30152; Physics
Frontier Center/JINA Center for the Evolution of the Elements (JINA-CEE), awarded by the US National Science Foundation. SRM thanks NSF grant AST-1616636. We would like to thank
M. Haywood and V. Hill for their  contribution and discussion. We want to thank the anonymous referee for her/his  extremly useful comments.

Funding for the Sloan Digital Sky Survey IV has been provided by the Alfred P. Sloan Foundation, the U.S. Department of Energy Office of
Science, and the Participating Institutions. SDSS acknowledges support and resources from the Center for High-Performance Computing at
the University of Utah. The SDSS web site is www.sdss.org.

SDSS is managed by the Astrophysical Research Consortium for the Participating Institutions of the SDSS Collaboration including the Brazilian Participation Group, the Carnegie Institution for Science, Carnegie Mellon University, the Chilean Participation Group, the French Participation Group, Harvard-Smithsonian Center for Astrophysics, Instituto de Astrofísica de Canarias, The Johns Hopkins University, Kavli Institute for the Physics and Mathematics of the Universe (IPMU) / University of Tokyo, Lawrence Berkeley National Laboratory, Leibniz Institut für Astrophysik Potsdam (AIP), Max-Planck-Institut für Astronomie (MPIA Heidelberg), Max-Planck-Institut für Astrophysik (MPA Garching), Max-Planck-Institut für Extraterrestrische Physik (MPE), National Astronomical Observatory of China, New Mexico State University, New York University, University of Notre Dame, Observatório Nacional / MCTI, The Ohio State University, Pennsylvania State University, Shanghai Astronomical Observatory, United Kingdom Participation Group, Universidad Nacional Autónoma de México, University of Arizona, University of Colorado Boulder, University of Oxford, University of Portsmouth, University of Utah, University of Virginia, University of Washington, University of Wisconsin, Vanderbilt University, and Yale University.
\end{acknowledgements}

\bibliographystyle{aa}
\bibliography{baadewindows_revised2}

\end{document}